\documentclass[aps,superscriptaddress,nofootinbib]{revtex4-2}

\usepackage{t1enc}
\usepackage[utf8]{inputenc}
\usepackage[toc,page]{appendix}
\usepackage{caption}
\usepackage{amsmath,amsfonts,amssymb,xfrac,nicefrac,esint,float}
\usepackage{xcolor}

\newcommand\nc{\newcommand}
\nc{\todo}[1]{}
\renewcommand{\todo}[1]{\textcolor{blue}{\texttt{[:~#1~:]}}}
\nc{\todoo}[1]{}
\renewcommand{\todoo}[1]{\textcolor{red}{\texttt{[:~#1~:]}}}
\nc{\SzM}[1]{\todoo{#1 -- \emph{SzM}}}
\nc{\vp}[1]{\todo{#1 -- \emph{VP}}}

\nc\lat\textit
\nc\ie{\lat{i.e.,\ }} \nc\etal{\lat{et al.\ }} \nc\etc{\lat{etc.\ }}
\nc\eg{\lat{e.g.,\ }} \nc\insitu{\lat{in situ}} \nc\QED{\lat{Q.E.D.}}
\nc\cf{cf.\ } \nc\wrt{w.r.t.\ }
\nc\adden{extra energy}
\nc\difficoeff{A}

\def\re#1{(\ref{#1})}

\nc\ttensor\mathbf
\nc\Tensor\boldsymbol
\nc\ident{\mathbf1}
\nc\zero{\ttensor{0}}

\nc\dd{{\rm d}}
\nc\pd{\partial}
\nc\pdt[1]{\frac{\pd #1}{\pd t}}
\nc\pder[2]{\frac{\pd #1}{\pd #2}}
\nc\pderr[3]{\left. \frac{\pd #1}{\pd #2} \right|_{#3}}
\nc\ppder[2]{\frac{\pd #1}{\pd \left( #2 \right) }}
\nc\ppderr[3]{\left. \frac{\pd #1}{\pd \left( #2 \right) } \right|_{#3}}
\nc\qdot[1]{\left( #1 \right) \dot{\vphantom{\left( #1 \right)}}}

\nc\qrho{\varrho}
\nc\qphi{\varphi}

\nc\qJE{J_E}
\nc\qqJE{\ttensor{J}_E}
\nc\qJU{J_U}
\nc\qqJU{\ttensor{J}_U}
\nc\qJS{J_S}
\nc\qqJS{\ttensor{J}_S}

\nc\he{\varepsilon}
\nc\qqh{{\ttensor{h}}}
\nc\qP{{\mathsf{P}}}
\nc\qqP{{\ttensor{P}}}
\nc\qqPi{{\Tensor{\Pi}}}
\nc\qqq{{\ttensor{q}}}
\nc\qqr{{\ttensor{r}}}
\nc\qv{{\mathsf{v}}}
\nc\qqv{{\ttensor{v}}}
\nc\qm{{\mathsf{m}}}
\nc\qM{{\mathsf{M}}}

\nc\bbS{\mathbb{S}}
\nc\cB{\mathcal{B}}

\nc\be{\Gamma_{\hskip -0.5ex e}}
\nc\br{\Gamma_{\hskip -0.5ex\qrho}}
\nc\bgr[1]{\Gamma_{\hskip -0.5ex \nabla \qrho}^{#1}}
\nc\bbgr{\Tensor{\Gamma}_{\hskip -0.5ex \nabla \qrho}}
\nc\bp{\Gamma_{\hskip -0.5ex \qphi}}
\nc\bgp[1]{\Gamma_{\hskip -0.5ex \nabla \qphi}^{#1}}
\nc\bbgp{\Tensor{\Gamma}_{\hskip -0.5ex \nabla \qphi}}
\nc\bv[1]{{{\Gamma_{\hskip -0.5ex{\qqv}}}^{}}^{}_{#1}}
\nc\bbv{{{\Tensor{\Gamma}_{\hskip -0.5ex{\qqv}}}^{}}^{}}

\nc\qf{_{\rm fl}}
\nc\id{_{\rm per}}

\nc\ee{{\rm e}}
\nc\ii{{\rm i}}

\nc\qS{\mathsf{S}}

\begin{document}
	
\title{Thermodynamic compatibility, holographic property and wave-function representation of generalized weakly nonlocal self-gravitating non-relativistic fluids}

\author{M\'aty\'as Sz\"ucs}
\affiliation{Department of Energy Engineering, Faculty of Mechanical Engineering, Budapest University of Technology and Economics, M\H{u}egyetem rkp.\ 3., H-1111 Budapest, Hungary}
\affiliation{Department of Theoretical Physics, HUN-REN Wigner Research Centre for Physics, Institute for Particle and Nuclear Physics, Konkoly Thege Mikl\'os St.\ 29--33, H-1121 Budapest, Hungary}
\affiliation{Montavid Thermodynamic Research Group, Society for the Unity of Science and Technology, Budapest, Hungary}

\author{P\'eter V\'an}
\email[]{van.peter@wigner.hun-ren.hu}
\affiliation{Department of Theoretical Physics, HUN-REN Wigner Research Centre for Physics, Institute for Particle and Nuclear Physics, Konkoly Thege Mikl\'os St.\ 29--33, H-1121 Budapest, Hungary}
\affiliation{Department of Energy Engineering, Faculty of Mechanical Engineering, Budapest University of Technology and Economics, M\H{u}egyetem rkp.\ 3., H-1111 Budapest, Hungary}
\affiliation{Montavid Thermodynamic Research Group, Society for the Unity of Science and Technology, Budapest, Hungary}

\date{\today}

\begin{abstract}

        Two pillars of physics---gravity and quantum mechanics---are usually formulated in entirely separate languages. Surprisingly, both can be expressed in the language of a fluid: a self-gravitating fluid for Newtonian gravity, and a quantum-pressure-carrying fluid for the Schr\"odinger equation. In this work we show that this resemblance {can be understood within a common framework based on how the Second Law of Thermodynamics constrains the class of non-dissipative continua considered here}. Using {a transparent thermodynamic procedure}, we derive the most general fluid model {within a chosen constitutive state space involving density, velocity, energy, an internal scalar field, and their spatial gradients}. {The framework recovers}, as special cases, the Euler equations, {a restricted family of Korteweg fluids} (including the hydrodynamic form of the Schr\"odinger and Gross--Pitaevskii equations), Newtonian self-gravitating fluids, and the Schr\"odinger--Newton system. {We identify the conditions under which perfect fluids exhibit \emph{classical holography}: the pressure-tensor divergence can then be represented by the gradient of a scalar potential.}
        {Classical holography yields a vorticity transport law that preserves sufficiently regular, initially irrotational flows and, under additional assumptions, enables a single complex-valued Schr\"odinger-like representation.} {While gravitational and quantum-hydrodynamic theories were investigated separately in our previous works, the present study develops a substantially more general scalar-internal-variable construction, refines its perfect-fluid analysis, and systematically embeds the earlier models within a unified framework.} The framework also {identifies a thermodynamically admissible density--gravity coupling that connects classical and quantum self-gravitating fluids, although its stability, and observational consequences remain to be explored.}

\end{abstract}
        

\maketitle

\section{Introduction}\label{sec:intro}

Fluids are everywhere: in the atmosphere, in stars and galaxies, inside living organisms, and at the heart of many engineering applications. Their motion is described almost universally by the Navier--Stokes---Fourier equations, a framework that has remained extraordinarily successful for nearly two centuries. Yet whenever physics has reached into a new regime---superfluids, Bose--Einstein condensates, capillary phenomena, self-gravitating gas clouds, or the hydrodynamic reformulation of quantum mechanics---the classical fluid equations had to be \emph{generalized}, often by adding higher-order spatial derivatives or new internal fields whose origin was not always transparent. The question this paper addresses is simple: \emph{is there a single, principled way to obtain such generalizations, and what do the resulting equations have in common?}

Our answer rests on a single guiding rule: the Second Law of Thermodynamics. The Second Law alone, applied carefully to a fluid whose state depends on density, velocity, energy, an internal scalar field and the gradients of all of these, fixes the most general form of the pressure tensor, the energy current and the evolution of the internal field {under the constitutive and modeling assumptions specified below}. Once this is done, three striking results emerge. First, {we identify the conditions under which non-dissipative (``perfect'') fluids of this class exhibit} a common geometric property that we term \emph{classical holography}: the divergence of the pressure tensor {can then be represented by} the gradient of a scalar potential, so that the {continuum momentum balance admits an equivalent} particle-like Newton equation. Second, when the flow is irrotational, the entire dynamics can be repackaged into a single complex-valued equation that has the structure of a Schr\"odinger equation and a {subclass} of Korteweg fluids and the Gross--Pitaevskii model emerge. {The resulting vorticity transport law also preserves sufficiently regular, initially irrotational flows, which is an important result in its own right.} Third, the internal scalar field {can be identified with} the gravitational potential; therefore, the same machinery reproduces Newtonian self-gravitating fluids and the Schr\"odinger--Newton system, and predicts a previously unrecognized, {thermodynamically admissible} density--gravity coupling{, whose physical viability remains to be established}.

The Navier--Stokes---Fourier equations provide a fundamental basis for modeling many physical phenomena, from classical to astrophysical fluid dynamics. Despite its wide range of applications, several generalizations of the theory are known, including non-Newtonian and viscoelastic fluids, fluids with internal spin, with nonlocal and higher-order terms or relativistic fluids.

Nonequilibrium thermodynamics, the theory of dissipative phenomena, serves as a universal tool for such generalizations. The ultimate challenge of the field is to characterize the constraints on the evolution of thermodynamic state variables and derive constitutive relations that fulfill the Second Law of Thermodynamics.

The history of nonequilibrium thermodynamics began nearly a century ago. The Norwegian-born physical chemist Lars Onsager is considered the father of the discipline. Based on statistical physical considerations, he established the reciprocity relations between thermodynamic forces and fluxes, which are now associated with his name \cite{Ons31a1,Ons31a2}. 

A decade later, Carl Eckart was the first, who derived the classical constitutive functions of thermofluid dynamics, \ie Fourier's law of heat conduction, Newton's law of viscosity and Fick's law of diffusion on the basis of macroscopic continuum theory without any statistical, microscopic background, relying solely on the application of balance equations and entropy inequality \cite{Eck40a1,Eck40a2}. The emerging theory---the so-called Classical Irreversible Thermodynamics---provides a uniform but heuristic framework for obtaining constitutive functions and is also suitable for chemical thermodynamics \cite{GroMaz62b}. 

However, compatibility with continuum mechanics required improvement of the heuristic background. Therefore, more rigorous methods have been developed for the evaluation of constrained inequalities, such as the Coleman-Noll and Liu procedures \cite{ColNol63a,Liu72a}. {Modeling} inertial effects and systems out of local equilibrium require a different method, where dissipative fluxes are introduced as independent thermodynamic state variables, and their evolution equations are constructed by thermodynamics \cite{Gya77a,JouAta92b,MulRug98b}. Generalization of this idea is the method of internal variables, when new fields---the so-called internal variables---are introduced in a thermodynamically consistent way \cite{ColGur67a,Ver97b,BerVan17b}. Some elements of the above-mentioned methodologies turned out to be predictive for unusual dissipative situations and modeling complex material behavior, such as heat conduction beyond Fourier's law, continuum mechanics beyond elasticity (viscoelasticity, rheology of fluids and solids, plasticity, damage mechanics, \dots), electric and magnetic relaxation, \etc \cite{MauMus94a1,MauMus94a2,VanAta08a,AssEta15a,KovVan15a}. Weakly nonlocal theories, where gradients of the thermodynamic state variables are also introduced in the thermodynamic state space, require the combination and improvement of previous ideas \cite{Van05a,Cim07a,VanAta08a}. This latter theory enables the analysis of the thermodynamic compatibility of field theories.

The main difficulty in thermodynamic modeling is the distinction between the emergent and fundamental aspects. In nonequilibrium thermodynamics the Second Law is fundamental. This is both a conceptual and a technical aspect, because without an explicit microscopic background, the constructive use of thermodynamic principles requires particular mathematical methods. Moreover, as long as only the general, objective elements are preserved and the particular, subjective, and statistical parts of the theory are properly removed, the fundamental aspect, the universality of thermodynamics is evident. Therefore, the methodology of nonequilibrium thermodynamics is simple and constructive, which nowadays goes far beyond the original visions of finding thermodynamically compatible constitutive functions. The Second Law of Thermodynamics appears to be the general principle for the construction of evolution equations. It can replace variational principles in ideal continua and derive dissipative evolution equations for non-ideal ones \cite{VanKov20a}. 

Classical holographic property, \ie when the surface traction of any closed surface can be equivalently transformed into a bulk force, seems to be a general property of perfect (\ie non-dissipative) continua. In other words, according to the classical holographic property the divergence of the pressure tensor of a perfect fluid can be expressed by a volumetric force density. At this time, the equation of motion can be formulated as
\begin{align}
	\qrho \dot \qqv + \nabla \cdot \qqP \qquad = \qquad \qrho (\dot \qqv + \nabla \Phi) \qquad = \qquad 0,
	\label{qholodem}\end{align}
where $\qrho$ is the {(mass)} density, $\qqv$ is the velocity, $\qqP$ is the pressure tensor and $\Phi$ is a (scalar) potential field; $\nabla$ and $\nabla\cdot$ are the gradient and divergence differential operators and
\begin{align}
	\label{eq:Dt}
	\dot \qqv = \left( \pdt{} + \qqv \cdot \nabla \right) \qqv
\end{align}
is the comoving or substantial time derivative (presented on velocity $\qqv$). {Note that the term ``classical holography'' is used here in a restricted continuum-mechanical sense and does not refer to the holographic principle of quantum gravity or to gauge/gravity dualities such as the AdS/CFT correspondence. The formal similarity lies in the dimensional reduction and boundary dependence, where a bulk force density can be equivalently represented through surface data. However, unlike quantum field-theoretic holography---where a higher-dimensional spacetime and gravity emerge non-locally from a lower-dimensional boundary theory---the classical counterpart operates within the same physical space. It refers to the equivalence between the surface traction representation associated with the pressure tensor and a bulk-force representation generated by a scalar potential. The terminology is closer in spirit to Gábor's original concept of holography, in which the information carried by an optical wavefront is encoded in a different representation from which the original wavefront can be reconstructed. Similarly, in the present continuum-mechanical setting, the force represented by the divergence of the pressure tensor is equivalently encoded in the reduced description provided by a single scalar potential.}

Equation \re{qholodem} is remarkable because it represents the transition between field and mass point representation of a physical system: classical holographic continua can be modeled in both theoretical frameworks. Classical holographic property appears in different forms in physics, mostly as a hidden aspect of particular, seemingly independent theories of perfect fields or continua, for example,
\begin{itemize}
	\item in the Friedmann equation, where the chemical potential is a mechanical potential as well, due to the Gibbs--Duhem relation \cite{Fri22b};
	\item in Newtonian gravity, where the effect of the gravitational potential can be represented via the divergence of a pressure tensor as a consequence of the Poisson equation \cite{VanAbe22a};
	\item in electrodynamics, it is well known, as part of the energy-momentum balance. However, the correct interpretation is dubious \cite{Mat24a}; 
	\item in Korteweg fluids, and therefore, in the fluid model of quantum mechanics through the quantum potential---a.k.a. Bohm potential---and the quantum pressure tensor as a consequence of the Bernoulli equation \cite{Boh52,Hol93b}.
\end{itemize}

Gravity and quantum phenomena are both universal, connecting them in a universal thermodynamic framework is natural. The connection between gravity and thermodynamics has long been known and is an active area of research. Dissipative quantum phenomena, with or without a thermodynamic framework have also been well developed. In both cases, the starting points are theories of gravity, in particular, general relativity \cite{Jac95a,Ver11a,Pad14a,EliEta06a}, or quantum mechanics \cite{Opp23a}. {A third route should be mentioned, closer in spirit to the present one. In entropic dynamics, quantum theory is derived as an application of entropic methods of inference: there is no underlying action principle, the dynamics being driven by entropy subject to suitable constraints, and the particular form of the quantum potential that leads to the Schr\"odinger equation follows from information geometry, that is, from the Fisher--Rao metric on the statistical manifold~\cite{Cat14a,CatEta14a,Cat19a}. The convergence with the present work is notable: two entirely different starting points---entropic inference on a statistical manifold, and the Second Law applied to a weakly nonlocal continuum---yield the same Fisher-information structure as the origin of the quantum potential, and both dispense with a variational principle. The differences are equally instructive, since entropic dynamics is non-dissipative by construction, whereas here dissipation is primary and perfect fluids arise as the zero-dissipation limit. Therefore, thermodynamics is not enforced in quantum or geometrical concepts and is not a special} appendix of \textit{ideal} gravitational or quantum phenomena. In thermodynamics, especially in nonequilibrium thermodynamics, \textit{dissipative} systems are general, and ideal systems are special cases of them with zero dissipation. The field equations, including the ideal equations, are restricted by the Second Law.

A possible verification of a general method is its compatibility with well-established theories. Newtonian gravity and basic quantum mechanics can be obtained using thermodynamic methodology \cite{VanAbe22a,AbeVan22a,Van23a}. In the first case, a single-component fluid with a weakly nonlocal scalar internal variable is investigated. The thermodynamical derivation resulted in notable extensions and generalizations; see for example, \cite{PszoVan24a}, and as a special case, when the evolution equation of the scalar field is not dissipative, the derivation leads to Newtonian gravity \cite{VanAbe22a,AbeVan22a}. In the second case, a single-component fluid with a weakly nonlocal extension in the density is considered, which resulted in a thermodynamically {restricted} family of Korteweg fluids. A detailed analysis has shown the conditions under which the Korteweg fluid evolution equations can be interpreted through a complex scalar field and become equivalent to superfluid theories (\eg Gross--Pitaevskii equation) and simplified to a probabilistic fluid and become equivalent to the Schrödinger equation \cite{Van23a}.

In this study, we investigate a combined situation: a single-component fluid with weak nonlocality in the density together with a weakly nonlocal scalar internal variable. We provide an explanation of the conditions for the classical holographic property, which in our approach {emerges, under the specified conditions, as a consequence} of the Second Law of Thermodynamics for perfect continua. A consequence of the holographic property is {a vorticity transport law that preserves initially irrotational flows under appropriate regularity conditions}. As a special case, when a perfect fluid is free of rotation, the balance of linear momentum can be reformulated in a Bernoulli equation for a velocity potential. In this case, the time evolution of the continuum is described via the mass continuity equation and the former Bernoulli equation, which---via the inverse Madelung transformation---can be represented in a single equation on a complex wave function. Then a natural combination of the fluid fields {leads to the the Schr\"odinger equation.} When the internal variable is identified with the gravitational potential, time evolution of self-gravitating fluids and as a particular result the Schr\"odinger--Poisson system is obtained. Therefore, the recent nonequilibrium thermodynamic treatment unites and explains seemingly independent observations and explores the gravity-quantum-thermodynamics connections.

{
\subsection{Main results and outline of the paper}

The main result of the present work is that several continuum models usually derived from distinct physical starting points can be recovered within a common thermodynamic framework: under the assumptions specified below, ordinary fluid dynamics, Newtonian self-gravitating fluids, and the hydrodynamic formulations of certain quantum models emerge as special cases of the same thermodynamically consistent continuum framework. In the dissipation-free, perfect-fluid realizations of these models, their common thermodynamic origin manifests itself in the classical holographic property: under the specific conditions identified below, the surface traction can be equivalently represented by a bulk force derived from a scalar potential.

In Sec.~\ref{sec:der}, a thermodynamically compatible family of weakly nonlocal fluids is derived via the Liu procedure. The resulting entropy inequality is then analyzed, and its simplest linear solution is presented. Although this section contains the principal technical derivation, it is essential for establishing the constitutive structure of the investigated fluid model, distinguishing its reversible and dissipative components, and identifying the conditions underlying the subsequent perfect-fluid analysis.

Readers interested mainly in the physical consequences may wish to focus on Sections~\ref{sec:per-flu} and \ref{sec:spec}. In Sec.~\ref{sec:per-flu}, the perfect-fluid equations are analyzed in general, the conditions for classical holography are established, and its consequences for preservation of initial irrotationality and the wave-function representation are investigated. Then, in Sec.~\ref{sec:spec}, several special cases are treated, including thermodynamically compatible Korteweg fluids, Newtonian gravity, and their combination, the Schrödinger--Poisson system. Furthermore, some straightforward generalizations of these models are examined. In particular, the generalized fluid model introduced here suggests a thermodynamically admissible density--gravity gradient coupling as a possible extension whose physical relevance remains to be established.
}

\section{Thermodynamically compatible family of weakly nonlocal fluids} 
\label{sec:der}

Let us investigate the thermo-mechanical processes of a single-component fluid in the non-relativistic spacetime. The advantage of a spacetime based approach is the covariant, reference frame independent formulation. Rigorous reference frame free treatment of non-relativistic fluid mechanics clarifies several concepts, that remain hidden in the usual frame dependent treatments. For example, the relationship between conductive and convective currents is a Galilean transformation; the fundamental balance equations are four divergences of the spacetime vector fields; and, the energy-momentum tensor is more sophisticated than in special relativity, \ie the covariant concept of energy still exists, but only as part of a third-order four-tensor \cite{Mat86a,Van17a}. A further important observation is that the spatial derivatives, \ie the gradients, are legitimate, objective constitutive variables; they are spacelike covectors, that do not transform when the reference frame changes. In addition, the entropy current density is a four-vector, therefore, if its timelike part, the entropy density, is a constitutive quantity, then its spacelike part, the entropy current density, must also be a constitutive quantity. Although the reference frame free treatment of fluid dynamics is well established, for ease of understanding and interpretation, we are using the usual relative quantities instead of any four-dimensional ones, but with the Galilean relativistic spacetime awareness.

The non-relativistic spacetime is considered flat and modeled by a four dimensional affine space. Time is absolute and the space of an inertial observer is a three dimensional oriented affine space provided by a Euclidean structure, \ie there exists a metric $ \qqh $, which identifies space-like vectors and covectors \cite{Mat20b}. In contrast to special relativistic space time, this Euclidean structure is interpreted only on the flat three dimensional instantaneous hyperspaces. In the following invariant notation is applied, however, technically complicated calculations are performed in (abstract) index notation, when vector and tensor indices are denoted by $ i, j, k, \hdots = 1, \dots, 3 $ and Einstein's summation convention is used. For instance, for the arbitrary vector $ \ttensor{a} $ the relationships $ a_i = h_{ij} a^j $ and $ a^i = \left( h^{-1} \right)^{ij} a_j \equiv h^{ij} a_j $ hold, thus $ h^{ij} h_{jk} = \delta^i_k $ is the Kronecker symbol. The metric is no longer displayed, its effect is simply represented by the position of the indices, \eg $ a^i h_{ij} a^j \equiv a_i a^i $. 

Eulerian description is applied, \ie field quantities are parametrized by instantaneous time $ t $ and spatial coordinates $ \qqr $ w.r.t.~an arbitrarily chosen reference frame. Assuming a non-polar fluid (\ie there exist no internal rotational degrees of freedom) the state of motion of a fluid element w.r.t.~the inertial observer is described via its velocity $ \qqv $ and thermodynamic state of the fluid element can be characterized via its density $ \qrho $ and (mass) specific internal energy $ u $. Possible processes of the fluid fulfill the conservations of mass, linear momentum and total energy, which are manifested in the balance equations and expressed as
\begin{align}
	\label{eq:bal-m}
	\dot \qrho + \qrho \nabla \cdot \qqv &= 0, \\
	\label{eq:bal-v}
	\qrho \dot \qqv + \nabla \cdot \qqP &= 0, \\
	\label{eq:bal-e}
	\qrho \dot e + \nabla \cdot \qqJE &= 0,
\end{align}
where 
\begin{align}
	\label{eq:tot-int-kin}
	e = u + \frac{1}{2} \qqv \cdot \qqv    
\end{align}
denotes the specific total energy, $ \qqP $ is the pressure tensor, which is symmetric as a consequence of conservation of angular momentum and $ \qqJE $ is the conductive current density of the total energy (as a reminder, $ \dot{ \ } $ is the substantial time derivative, given already in \re{eq:Dt} and $ \nabla $ is the nabla operator). 

The fundamental balances determine the dynamics of the elements of the \emph{state space}, which is spanned by the state variables $ \qrho $, $ \qqv $ and $ e $. If the constitutive functions $ \qqP $ and $ \qqJE $ are given functions of the state space, then for given initial and boundary conditions the solution of the evolution equations \re{eq:bal-m}--\re{eq:bal-e} in a fixed spatial point is a curve in the state space, parameterized by the absolute time.

Only minimal requirements need to be imposed to achieve the most general form of the evolution equations of the state variables (within the selected state space). Therefore, we only expect that the constitutive functions $ \qqP $ and $ \qqJE $ should be thermodynamically compatible, \ie these are restricted only by the Second Law of Thermodynamics, which imposes constraints on the form of the constitutive functions. The Second Law of Thermodynamics claims that entropy increases in any insulated system, therefore, the source term of the local entropy balance, \ie the entropy production rate density, must be nonnegative, which is expressed by the entropy inequality
\begin{align}
	\label{eq:s-ineq}
	0 \le \qrho \dot s + \nabla \cdot \qqJS,
\end{align}
where, the specific entropy function $ s $ and the entropy current density $ \qqJS $ are also constitutive functions, which have to be determined. Accordingly, the realized process has to satisfy both the evolution equations \re{eq:bal-m}--\re{eq:bal-e} and the entropy inequality \re{eq:s-ineq}. From a mathematical point of view, \re{eq:s-ineq} is a conditional inequality, since entropy increase along the process, which is governed by the dynamical system prescribed on the state variables \re{eq:bal-m}--\re{eq:bal-e}.

The classical method to solve the conditional inequality is the heuristic \emph{divergence separation}. Accordingly, time evolution of the specific entropy is calculated in terms of the elements of the state space, and the constraints are then substituted into this form. Then the entropy current density is---heuristically---identified through recognizing total divergence terms to obtain an interpretable and solvable quadratic flux-force expression for the entropy production rate density. Subsequently, quasi-linear constitutive functions among fluxes and forces can be identified \cite{GroMaz62b}. The advantage of the method is also itself's disadvantage, the calculations rely mostly on physical intuitions rather than on mathematical foundations.

The insight and scope of a heuristic approach can be improved and extended by rigour. Then it is applicable for more complex systems. Such rigorous techniques are the \emph{Coleman--Noll procedure} \cite{ColNol63a} and the \emph{Liu procedure} \cite{Liu72a}. Both methods first fix the \emph{constitutive state space}, \ie the variables, which define the constitutive functions. The constitutive state space may include the derivatives of the state space, for example, in the case of Newtonian fluids the viscous part of the pressure tensor is proportional to the velocity gradient \cite{Gya70b} or in the case of Korteweg fluids, the reversible part of the pressure tensor containing terms connecting to the gradient and the Hessian of density \cite{Kor901a}. Therefore, the heuristic aspects are removed by distinguishing between constitutive functions and their variables. While the Coleman--Noll procedure investigates the possible solutions of the Clausius--Duhem inequality---an alternative of the entropy inequality in the energy dimension---, the Liu procedure directly evaluates the entropy inequality. The basic idea behind the Coleman--Noll method is to expand the entropy inequality in terms of the elements of the constitutive state space and replace the evolution equations into it. Coefficients of the terms with undetermined signs (typically time derivatives) have to be zero to ensure the increase of entropy also in the most general case, thus, only the positive semi-definiteness of the remaining parts has to be ensured. The more sophisticated Liu procedure generalizes this idea, namely, entropy inequality as a conditional inequality is solved via using so-called Lagrange--Farkas multipliers (a generalization of classical Lagrange multipliers applied for conditional extremum problems). The constraints multiplied by the initially unknown Lagrange--Farkas multipliers are subtracted from the entropy inequality, which inequality is then solved for the multipliers via Liu's theorem \cite{Liu72a,HauKir02a}, a special case of the Farkas' lemma \cite{TriAta08a}. This latter method enables the direct determination of not only the entropy production rate density but also the entropy current density in terms of the constitutive state space. In what follows, we present the thermodynamic derivation of a weakly nonlocal generalized fluid model, applying the Liu procedure. Theoretically, no differences in the outcome are expected whether the Coleman--Noll or the Liu procedure is applied, it is only a question of convenience which to choose. An application of the Coleman--Noll procedure for weakly nonlocal continua is presented in \cite{Pao22a,Pao23ac,Pao25}.

\subsection{Defining the constitutive state space}

Let us now suppose that dynamics of the fluid cannot be characterized only via equations \re{eq:bal-m}--\re{eq:bal-e}, but an additional scalar field denoted by $ \qphi $ also contributes to the dynamics. {Such a degree of freedom need not be directly observable or independently measurable; its presence may instead be inferred from constitutive behavior that cannot be described by the conventional hydrodynamic variables alone, for example, from additional contributions to the pressure or stored energy, or from relaxation effects. In many applications, an internal variable can be eliminated from the coupled evolution equations, in which case its effects reappear as a generalized constitutive relation between directly interpretable physical quantities, typically through time derivatives, relaxation terms, or memory effects.} From a thermodynamical point of view this field is an \textit{internal variable}. {In addition to describing constitutive effects of material origin, internal variables may also represent field-matter interactions, as discussed in Sec.~\ref{sec:spec}. At this stage, no particular physical interpretation is assigned to it; its specific meaning is determined only through the subsequent constitutive assumptions.} The evolution equation of $ \qphi $ is written in a general form as
\begin{align}
	\label{eq:xi-evol}
	\dot \qphi + f = 0,
\end{align}
where $ f $---similarly to $ \qqP $ and $ \qqJE $---is a constitutive function. Therefore, the state space is extended, it is now spanned by the fields $ \left( \qrho, \qqv, e, \qphi \right) $. We rely on our previous assumption, \ie the most general time evolution of the internal variable is determined only and exclusively by the Second Law of Thermodynamics.

As we have previously mentioned, classical constitutive equations, such as Fourier's heat conduction law or Newton's viscosity law, are gradient-dependent. The former presents a quasi-linear relationship between the heat current density and the temperature gradient, while the latter one between the viscous pressure and the velocity gradient. Therefore, we can deduce that the above introduced constitutive functions---$ s $, $ \qqJS $, $ \qqP $, $ \qqJE $ and $ f $--- depend not only on the state variables $ \left( \qrho, \qqv, e, \qphi \right) $, but can depend on the gradients of the state variables, too. Weakly nonlocal theories with gradient-dependent state spaces and thermodynamic potentials are classified according to the highest order spatial derivatives among the constitutive variables. 

The choice of the constitutive state space is our most important \emph{assumption}. Now, it is assumed to be second-order weakly nonlocal in the density and in the internal variable and first-order weakly nonlocal in the velocity and in the energy, \ie
\begin{align}
	\label{eq:const-st-sp}
	\Big( \qrho, \nabla \qrho, \nabla \otimes \nabla \qrho, \qqv, \nabla \qqv, e, \nabla e, \qphi, \nabla \qphi, \nabla \otimes \nabla \qphi \Big).
\end{align}
Without the internal variable and with a first-order constitutive state space in every state variable, one obtains the Navier--Stokes---Fourier equations \cite{Szu22}. {Throughout the derivation, the constitutive functions and the corresponding fields are assumed to be sufficiently regular for the indicated spatial derivatives, functional derivatives, and integration-by-parts transformations to be well defined.}

\subsubsection{On thermostatics of weakly nonlocal continua}

Classical theory of thermodynamics is based on the concept of downscaling (\ie the transformation from bodies to continua), which seemingly prohibit the presence of nonlocal variables, such as the gradients of the state variables. 

Let $ X^a \ ( a = 1, \dots, A ) $ denote the minimum but sufficient number of extensive (proportional to the size of the system) thermodynamic state variables that uniquely characterize the dynamics of a body. The body's entropy $ S $ given in the variables $ X^a $ is a \emph{thermodynamic potential} function, \ie it condenses all information about the material's static behavior, which are inherited by the various equations of state. More precisely, the entropy of a thermodynamic body is the potential function of the (co)vector field spanned by the intensive thermodynamic state functions $ Y_a \ ( a = 1, \dots, A ) $, characterized by the Gibbs relation
\begin{align}
	\label{eq:Gibbs-S}
	\dd S \left( X^1, X^2, \dots, X^A \right) = \sum_{ a = 1 }^{ A } Y_a \dd X^a.
\end{align}
Correspondingly, the intensive state functions---usually manifested in the equations of state---are defined as partial derivatives of entropy, that is,
\begin{align}
	\label{eq:S-pder}
	Y_a \left( X^1, X^2, \dots, X^A \right) &= \left. \frac{ \pd S }{ \pd X^a } \right|_{ X^b, \ b = 1, \dots, A, \ a \neq b }, &
	a &= 1, \dots, A.
\end{align}
In the following, for the sake of easier transparency, we will use the notation $ \left. \frac{ \pd S }{ \pd X^a } \right|_{ X^b } = \left. \frac{ \pd S }{ \pd X^a } \right|_{ X^b, \ b = 1, \dots, A, \ a \neq b } $.

Entropy is also an extensive quantity, which is reflected by its first-order homogeneity in the variables $ X^a $. The consequence of first order Euler homogeneity is that for any arbitrary scalar extensive thermodynamic quantity $ X^{\tilde{a}} $ the $ X^{\tilde{a}} $-specific entropy can be introduced, which is a function of the corresponding specific quantities $ x^a = \frac{X^a}{X^{\tilde{a}}} $ (especially, if $ {\tilde{a}} = a $ then $ x^{\tilde{a}} \equiv 1 $). For instance, the $ X^1 $-specific entropy $ s_{X^1}^{} $ is defined through the relationship
\begin{align}
	\label{eq:S-hom}
	S \left( X^1, X^2, \dots, X^A \right) = X^1 s_{X^1}^{} \left( \frac{X^2}{X^1}, \dots, \frac{X^A}{X^1} \right) = X^1 s_{X^1}^{} \left( x^2, \dots, x^A \right).
\end{align}
Applying \re{eq:S-hom} to obtain the intensive state functions one finds
\begin{align}
	\nonumber
	Y_a \left( X^1, X^2, \dots, X^A \right) &= \left. \frac{ \pd S }{ \pd X^a } \right|_{ X^b } = \left. \frac{ \pd \left( X^1 s_{X^1}^{} \right) }{ \pd X^a } \right|_{ X^b } = X^1 \left. \frac{ \pd s_{X^1}^{} }{ \pd x^a } \right|_{ x^b } \cdot \frac{\pd x^a}{\pd X^a} \\
	\label{eq:S-pder-1}
	&= \left. \frac{ \pd s_{X^1}^{} }{ \pd x^a } \right|_{ x^b } = Y_a \left( x^2, \dots, x^A \right), \qquad
	a = 2, \dots, A,
\end{align}
and if $ a = 1 $ then 
\begin{align}
	\nonumber
	Y_1 \left( X^1, X^2, \dots, X^A \right) &= \left. \frac{ \pd S }{ \pd X^1 } \right|_{ X^b, \ b \neq = 1 } = \left. \frac{ \pd \left( X^1 s_{X^1}^{} \right) }{ \pd X^1 } \right|_{ X^b, \ b \neq = 1 } = s_{X^1}^{} \left( x^2, \dots, x^A \right) + X^1 \sum_{a=2}^A \left. \frac{ \pd s_{X^1}^{} }{ \pd x^a } \right|_{ x^b } \frac{\pd x^a}{\pd X^1} \\
	\label{eq:S-pder-2}
	& \stackrel{\re{eq:S-pder-1}}{=} s_{X^1}^{} \left( x^2, \dots, x^A \right) - \sum_{a=2}^A Y_a \left( x^2, \dots, x^A \right) x^a.
\end{align}
These observations highlight that the intensive state functions, as well as the $ X^1 $-specific entropy, are independent of the size of the body (characterized by $ X^1 $), therefore, the $ X^1 $-specific formulation applies to the material, not the body. Furthermore, multiplying \re{eq:S-pder-2} with $ X^1 $ and applying \re{eq:S-hom} one obtains a direct consequence of Euler homogeneity, the Euler relation:  
\begin{align}
	\label{Ms}
	S \left( X^1, X^2, \dots, X^A \right) = \sum_{ a = 1 }^{ A } Y_a \left( x^2, \dots, x^N \right) X^a.
\end{align}
Finally, forming the differential of \re{eq:S-hom} via $ \dd X^a = X^1 \dd x^a + x^a \dd X^1 $ and applying \re{eq:Gibbs-S} and \re{eq:S-pder-2}, the Gibbs relation for the $ X^1 $-specific entropy follows, \ie
\begin{align}
	\label{eq:Gibbs-s}
	\dd s_{X^1} \left( x^2, \dots, x^A \right) = \sum_{ a = 2 }^{ A } Y_a \left( x^2, \dots, x^N \right) \dd x^a.
\end{align}
The downscaling of physical properties based on the Euler homogeneity of the entropy (as well as of other thermodynamic potentials) opens the way towards the localization of thermodynamic theories. Assuming \emph{local thermodynamic equilibrium}, intensive state functions can be introduced through the specific entropy \re{eq:S-pder-1}. When local equilibrium is homogeneous, then the backward direction, \ie upscaling from local to global level is self-evident. However, this concept of extensivity is based on \emph{homogeneous} thermodynamic bodies. Any simple generalization, such as mixed order and fractional order Euler homogeneities (see, e.g., \cite{QueEta17a}) are the same from this point of view: homogeneous thermodynamic equilibrium is scaled down. 

To introduce the concept of extensivity for weakly nonlocal material models, the key aspect is the formal realization that local thermodynamic equilibrium is not necessarily homogeneous because of the presence of gradient variables in the state space. Then, the continuum theory must be considered as the starting point of thermodynamic modeling. Accordingly, at the local level, it is natural to assume that specific entropy depends on the gradients of physical quantities, however, the usual differential formalism of Gibbsian homogeneous thermodynamics with a suitable simple modification of the thermodynamic framework can be preserved. Upscaling is not straightforward, it can be realized through integration on the volume of the body $ \mathcal{V} $, \ie
\begin{align}
	S = \int\limits_{\mathcal{V}} \qrho s \dd V,
\end{align}
which reflects that for macroscopic thermodynamic bodies boundary conditions are also important, \ie thermodynamic equilibrium may depend on the shape of the body (\eg in elasticity and plasticity \cite{FleEta94a,Ber23b}).

All these highlight that the local Gibbs relation \re{eq:Gibbs-s} is more general than its global version \re{eq:Gibbs-S}. This extensivity reversal is a construction method for thermodynamic potentials in small-system thermodynamics \cite{Hil92a,BedEta23b}. In the case of weakly nonlocal continua and fields, the homogeneity of thermodynamic equilibrium is an exception; therefore, the Euler homogeneous upscaling of thermodynamic potentials and equations of state is not straightforward (but is possible in an effective sense, see, for example, for self-gravitating fluids in \cite{GioEta19a}).

\subsection{Solution of the entropy inequality}

The entropy inequality \re{eq:s-ineq} is constrained by the balances and evolution equations \re{eq:bal-m}--\re{eq:bal-e} and \re{eq:xi-evol}. To determine a solution of this constrained inequality we apply Liu's procedure, whose mathematical foundations rest on \textit{Liu's theorem}. Let $ \mathbb{V} $ and $ \mathbb{U} $ be finite dimensional vector spaces, their dual spaces are denoted by $ \mathbb{V}^* $ and $ \mathbb{U}^* $, furthermore, let $ \mathbf{a} \in \mathbb{V}^* $, $ b \in \mathbb{R} $, $ \mathbf{b} \in \mathbb{U} $, and $ \mathbf{A} : \mathbb{V} \to \mathbb{U} $ is a linear map. The inequality
\begin{align}
	\label{eq:Liu-th-1}
	\left ( \mathbf{a} | \mathbf{x} \right) + b \ge 0
\end{align}
holds for all $ \mathbf{x} \in \mathbb{V} $ such that
\begin{align}
	\label{eq:Liu-th-2}
	\mathbf{A} \mathbf{x} + \mathbf{b} = \mathbf{0}_{\mathbb{U}}^{},
\end{align}
if and only if there exists a so-called Lagrange--Farkas multiplier $ \boldsymbol{\Gamma} \in \mathbb{U}^* $ such that
\begin{align}
	\label{eq:Liu-th-3}
	\mathbf{a} - \mathbf{A}^*\boldsymbol{\Gamma} &= \mathbf{0}_{\mathbb{V}^*}, \\
	\label{eq:Liu-th-4}
	b - \left ( \boldsymbol{\Gamma} | \mathbf{b} \right) & \ge 0,
\end{align}
where $ \mathbf{A}^* : \mathbb{U}^* \to \mathbb{V}^* $ denotes the transpose of $ \mathbf{A} $. The assumptions of the algebraic Liu's theorem, \ie inequality \re{eq:Liu-th-1} and equation \re{eq:Liu-th-2} correspond to entropy inequality \re{eq:s-ineq} and evolution equations \re{eq:bal-m}--\re{eq:bal-e} and \re{eq:xi-evol}, respectively, and $ \mathbf{x} $ denote the elements of the so-called \emph{process direction space}, which is spanned by the elements of (both temporal and spatial) derivatives of the constitutive state space which are not included in the constitutive state space itself. For our investigated situation the process direction space is spanned by
\begin{align}
	\Big( \dot \qrho, \qdot{\nabla \qrho}, \qdot{\nabla \otimes \nabla \qrho}, \nabla \otimes ( \nabla \otimes \nabla \qrho ), \dot \qqv, \qdot{\nabla \qqv}, \nabla \otimes \nabla \qqv, \dot e, \qdot{\nabla e}, \nabla \otimes \nabla e, \dot \qphi, \qdot{\nabla \qphi}, \qdot{\nabla \otimes \nabla \qphi}, \nabla \otimes ( \nabla \otimes \nabla \qphi ) \Big).
\end{align}
Equation \re{eq:Liu-th-3} is called the Liu-equation, which allows to determine specific entropy and entropy current density as the functions of the elements of the process direction space, while inequality \re{eq:Liu-th-4} characterizes the dissipation. Liu-equations and dissipation inequality---analogously to conditional extrema problems---can be also expressed in the multiplier form 
\begin{align}
	\label{eq:multiplier-form}
	\left ( \mathbf{a} | \mathbf{x} \right) + b - \left ( \boldsymbol{\Gamma} | \mathbf{A} \mathbf{x} + \mathbf{b} \right) = \left( \mathbf{a} - \mathbf{A}^* \boldsymbol{\Gamma} | \mathbf{x} \right) + \left( b - \left ( \boldsymbol{\Gamma} | \mathbf{b} \right) \right) \ge 0,
\end{align}
which is preferred to use in calculations. The justification of using the algebraic theorem for differential equations requires that the elements of the process direction space have arbitrary values, which is ensured by the initial and boundary conditions of the system of evolution equations \cite{MusEhr96a,HauKir02a}.

In the case of higher-order, weakly nonlocal state spaces, the balance and evolution equations alone do not provide a sufficient number of constraints for the application of Liu's procedure to give physically meaningful results. The reason for this is that the derivatives of some constraints may be considered as further constraints, consequently, the time evolution of the gradient variables also constrains the entropy inequality \cite{Van05a,Cim07a,Van09a1}. In our case the weakly nonlocal variables are $ \qrho $ and $ \qphi $ [see the Hessians of these quantities in \re{eq:const-st-sp}], thus the gradients of \re{eq:bal-m} and \re{eq:xi-evol}, \ie
\begin{align}
	\label{eq:grad-bal-m}
	\nabla \dot \qrho + \nabla \left( \qrho \nabla \cdot \qqv \right) &= 0, \\
	\label{eq:grad-xi-evol}
	\nabla \dot \qphi + \nabla f &= 0
\end{align}
are additional constraints. Let us call the attention that the substantial time derivative and the gradient do not commute,
\begin{align}
	\pd_i \dot \qphi = \pd_i \left( \pdt{\qphi} + \qv^j \pd_j \qphi \right) = \pdt{\left( \pd_i \qphi \right)} + \pd_i \qv^j \pd_j \qphi + \qv^j \pd_{ij} \qphi = \left( \pd_i \qphi \right) \dot{\vphantom{\left( \pd_i \qphi \right)}} + \pd_i \qv^j \pd_j \qphi
\end{align}
so,
\begin{align}
	\label{eq:dot-noncomm}
	\nabla \dot \qphi = 
	\qdot{ \nabla \qphi } + \nabla \qqv \cdot \nabla \qphi.
\end{align}
Therefore, \re{eq:grad-bal-m} and \re{eq:grad-xi-evol} are equivalent to
\begin{align}
	\label{eq:gradrho-evol}
	\qdot{ \nabla \qrho } + \nabla \qqv \cdot \nabla \qrho + \left( \nabla \cdot \qqv \right) \nabla \qrho + \qrho \nabla \left( \nabla \cdot \qqv \right) &= 0, \\
	\label{eq:gradxi-evol}
	\qdot{ \nabla \qphi } + \nabla \qqv \cdot \nabla \qphi + \nabla f &= 0,
\end{align}
which directly describe the time evolution of the gradient variables.

Accordingly, we have formulated the necessary additional constraints, thus the entropy inequality \re{eq:s-ineq} must be solved along the conditions provided by the balance equations \re{eq:bal-m}--\re{eq:bal-e} and time evolution equations \re{eq:xi-evol}, \re{eq:gradrho-evol} and \re{eq:gradxi-evol}. Applying the multiplier form [\cf \re{eq:multiplier-form}], our conditional inequality is
\begin{align}
	\nonumber
	0 &\le \qrho \dot s + \nabla \cdot \qqJS - \br \left( \dot \qrho + \qrho \nabla \cdot \qqv \right) - \bbgr \cdot \left[ \qdot{ \nabla \qrho } + \nabla \qqv \cdot \nabla \qrho + \left( \nabla \cdot \qqv \right) \nabla \qrho + \qrho \nabla \left( \nabla \cdot \qqv \right) \right] \\
	& \hskip 2.7ex - \bbv \cdot \left( \qrho \dot \qqv + \nabla \cdot \qqP \right) - \be \left( \qrho \dot e + \nabla \cdot \qqJE \right) - \bp \left( \dot \qphi + f \right) - \bbgp \cdot \left[ \qdot{ \nabla \qphi } + \nabla \qqv \cdot \nabla \qphi + \nabla f \right],
\end{align}%
{
Although the resulting expression is necessarily lengthy, we retain its complete systematic form because of the algorithmic nature of the Liu procedure. The constitutive functions are expanded in terms of the elements of the constitutive state space, and the resulting terms are grouped according to the independent elements of the process-direction space. At this intermediate stage, assigning separate physical interpretations to the individual terms would obscure the systematic structure of the procedure; their thermodynamic meaning becomes transparent only after the corresponding coefficients have been separated. Accordingly, the conditional entropy inequality is equivalent to
}
\begin{align}
	\nonumber
	0 &\le \left( \qrho \pder{s}{\qrho} - \br \right) \underline{\dot \qrho} + \left( \qrho \ppder{s}{\pd_i \qrho} - \bgr{i} \right) \underline{\qdot{\pd_i \qrho}} + \qrho \ppder{s}{\pd_{ij} \qrho} \underline{\qdot{\pd_{ij} \qrho}} + \left( \pder{s}{\qv^i} - \bv{i} \right) \qrho \underline{\dot \qv^i} + \qrho \ppder{s}{\pd_j \qv^i} \underline{\qdot{\pd_j \qv^i}} \\
	\nonumber
	& \hskip 2.7ex + \left( \pder{s}{e} - \be \right) \qrho \underline{\dot e} + \qrho \ppder{s}{\pd_i e} \underline{\qdot{\pd_i e}} + \left( \qrho \pder{s}{\qphi} - \bp \right) \underline{\dot \qphi} + \left( \qrho \ppder{s}{\pd_i \qphi} - \bgp{i} \right) \underline{\qdot{\pd_i \qphi}} + \qrho \ppder{s}{\pd_{ij} \qphi} \underline{\qdot{\pd_{ij} \qphi}} \\
	\nonumber
	& \hskip 2.7ex + \left( \pder{\qJS^j}{\qrho} - \bv{i} \pder{\qP^{ij}}{\qrho} - \be \pder{\qJE^j}{\qrho} - \bgp{j} \pder{f}{\qrho} \right) \pd_j \qrho + \left( \ppder{\qJS^j}{\pd_k \qrho} - \bv{i} \ppder{\qP^{ij}}{\pd_k \qrho} - \be \ppder{\qJE^j}{\pd_k \qrho} - \bgp{j} \ppder{f}{\pd_k \qrho} \right) \pd_{jk} \qrho \\
	\nonumber
	& \hskip 2.7ex + \left( \ppder{\qJS^j}{\pd_{kl} \qrho} - \bv{i} \ppder{\qP^{ij}}{\pd_{kl} \qrho} - \be \ppder{\qJE^j}{\pd_{kl} \qrho} - \bgp{j} \ppder{f}{\pd_{kl} \qrho} \right) \underline{\pd_{jkl} \qrho} + \left( \pder{\qJS^j}{\qv^i} - \bv{l} \pder{\qP^{lj}}{\qv^i} - \be \pder{\qJE^j}{\qv^i} - \bgp{j} \pder{f}{\qv^i} \right) \pd_j \qv^i \\
	\label{eq-s-ineq-nonloc}
	& \hskip 2.7ex + \left( \ppder{\qJS^j}{\pd_k \qv^i} - \frac{\qrho}{2} \bgr{l} \left( \delta_i^j \delta_l^k + \delta_i^k \delta_l^j \right) - \bv{l} \ppder{\qP^{lj}}{\pd_k \qv^i} - \be \ppder{\qJE^j}{\pd_k \qv^i} - \bgp{j} \ppder{f}{\pd_k \qv^i} \right) \underline{\pd_{jk} \qv^i} \\
	\nonumber
	& \hskip 2.7ex + \left( \pder{\qJS^j}{e} - \bv{i} \pder{\qP^{ij}}{e} - \be \pder{\qJE^j}{e} - \bgp{j} \pder{f}{e} \right) \pd_j e + \left( \ppder{\qJS^j}{\pd_k e} - \bv{i} \ppder{\qP^{ij}}{\pd_k e} - \be \ppder{\qJE^j}{\pd_k e} - \bgp{j} \ppder{f}{\pd_k e} \right) \underline{\pd_{jk} e} \\
	\nonumber
	& \hskip 2.7ex + \left( \pder{\qJS^j}{\qphi} - \bv{i} \pder{\qP^{ij}}{\qphi} - \be \pder{\qJE^j}{\qphi} - \bgp{j} \pder{f}{\qphi} \right) \pd_j \qphi + \left( \ppder{\qJS^j}{\pd_k \qphi} - \bv{i} \ppder{\qP^{ij}}{\pd_k \qphi} - \be \ppder{\qJE^j}{\pd_k \qphi} - \bgp{j} \ppder{f}{\pd_k \qphi} \right) \pd_{jk} \qphi \\
	\nonumber
	& \hskip 2.7ex + \left( \ppder{\qJS^j}{\pd_{kl} \qphi} - \bv{i} \ppder{\qP^{ij}}{\pd_{kl} \qphi} - \be \ppder{\qJE^j}{\pd_{kl} \qphi} - \bgp{j} \ppder{f}{\pd_{kl} \qphi} \right) \underline{\pd_{jkl} \qphi} \\
	\nonumber
	& \hskip 2.7 ex - \bp f - \left( \qrho \br \delta_j^i + \bgr{i} \pd_j \qrho + \bgp{i} \pd_j \qphi \right) \pd_i \qv^j - \bgr{i} \pd_i \qrho \pd_j \qv^j.
\end{align}
Here, and in the following, we do not denote separately the variables that are kept constant in partial differentiation, which is clearly indicated by introducing the corresponding functions. 

The coefficients of the elements of the process direction space [the underlined terms in \re{eq-s-ineq-nonloc}] {must vanish separately and yield the following Liu equations:}
\begin{align}
	\label{eq:Liu-1}
	\dot \qrho &: &
	\qrho \pder{s}{\qrho} &= \br, &
	\qdot{\pd_i \qrho} &: &
	\qrho \ppder{s}{\pd_i \qrho} &= \bgr{i}, &
	\qdot{\pd_{ij} \qrho} &: &
	\ppder{s}{\pd_{ij} \qrho} &= 0, \\
	\label{eq:Liu-4}
	\dot \qv^i &: &
	\pder{s}{\qv^i} &= \bv{i}, &
	\qdot{\pd_j \qv^i} &: &
	\ppder{s}{\pd_j \qv^i} &= 0, & & & &\\
	\label{eq:Liu-6}
	\dot e &: &
	\pder{s}{e} &= \be, &
	\qdot{\pd_i e} &: &
	\ppder{s}{\pd_i e} &= 0, & & & &\\
	\label{eq:Liu-8}
	\dot \qphi &: &
	\qrho \pder{s}{\qphi} &= \bp, &
	\qdot{\pd_i \qphi} &: &
	\qrho \ppder{s}{\pd_i \qphi} &= \bgp{i}, &
	\qdot{\pd_{ij} \qphi} &: &
	\ppder{s}{\pd_{ij} \qphi} &= 0,
\end{align}
\begin{align}
	\label{eq:Liu-11}
	\pd_{jkl} \qrho &: &
	\ppder{\qJS^j}{\pd_{kl} \qrho} &= \bv{i} \ppder{\qP^{ij}}{\pd_{kl} \qrho} + \be \ppder{\qJE^j}{\pd_{kl} \qrho} + \bgp{j} \ppder{f}{\pd_{kl} \qrho}, \\
	\label{eq:Liu-12}
	\pd_{jk} \qv^i &: &
	\ppder{\qJS^j}{\pd_k \qv^i} &= \bv{l} \ppder{\qP^{lj}}{\pd_k \qv^i} + \be \ppder{\qJE^j}{\pd_k \qv^i} + \bgp{j} \ppder{f}{\pd_k \qv^i} + \frac{\qrho}{2} \bgr{l} \left( \delta_i^j \delta_l^k + \delta_i^k \delta_l^j \right), \\
	\label{eq:Liu-13}
	\pd_{jk} e &: &
	\ppder{\qJS^j}{\pd_k e} &= \bv{i} \ppder{\qP^{ij}}{\pd_k e} + \be \ppder{\qJE^j}{\pd_k e} + \bgp{j} \ppder{f}{\pd_k e}, \\
	\label{eq:Liu-14}
	\pd_{jkl} \qphi &: &
	\ppder{\qJS^j}{\pd_{kl} \qphi} &= \bv{i} \ppder{\qP^{ij}}{\pd_{kl} \qphi} + \be \ppder{\qJE^j}{\pd_{kl} \qphi} + \bgp{j} \ppder{f}{\pd_{kl} \qphi}.
\end{align}
Note that in \re{eq:Liu-12} $ \qrho \bgr{i} \pd_{ij} \qv^j = \frac{\qrho}{2} \bgr{l} \left( \delta_i^j \delta_l^k + \delta_i^k \delta_l^j \right) \pd_{jk} \qv^i $, \ie the expression $ \bbgr \cdot \qrho \nabla \left( \nabla \cdot \qqv \right) $ appearing through \re{eq:gradrho-evol} is symmetrized, since its antisymmetric part is invisible in the dissipation inequality due to the symmetry of its coefficient. Therefore, we obtain a different solution for the pressure tensor than without the symmetry requirement \cite{AndEta98a,Mor23a,BhaEta24a,Van25a}.

According to equations \re{eq:Liu-1}--\re{eq:Liu-8} one can deduce that $ s = s \left( \qrho, \nabla \qrho, \qqv, e, \qphi, \nabla \qphi \right) $, \ie specific entropy depends on the gradient of density and of the internal variable, therefore, weakly nonlocal effects are also included in the thermostatic behavior of the material. The variables of the entropy function, $ \left( \qrho, \nabla \qrho, \qqv, e, \qphi, \nabla \qphi \right) $, determine the \emph{thermodynamic state space}. Furthermore, this means that the applied Lagrange--Farkas multipliers $ \br $, $ \bbgr $, $ \bbv $, $ \be $, $ \bp $ and $ \bbgp $ depend only on the variables $ \left( \qrho, \nabla \qrho, \qqv, e, \qphi, \nabla \qphi \right) $. Integrating equations \re{eq:Liu-11}--\re{eq:Liu-14} entropy current density is obtained, \ie
\begin{align}
	\label{eq:JS-nonloc-ind}
	\qJS^j = \pder{s}{\qv^i} \qP^{ij} + \pder{s}{e} \qJE^j + \qrho \ppder{s}{\pd_j \qphi} f + \frac{\qrho^2}{2} \ppder{s}{\pd_l \qrho} \left( \delta_l^j \delta_i^k + \delta_l^k \delta_i^j \right) \pd_k \qv^i + \mathcal{J}^j.
\end{align}
In the invariant notations reads as 
\begin{align}
	\label{eq:JS-nonloc}
	\qqJS = \pder{s}{\qqv} \cdot \qqP + \pder{s}{e} \qqJE + \qrho \ppder{s}{\nabla \qphi} f + \frac{\qrho^2}{2} \left[ \ppder{s}{\nabla \qrho} \left( \nabla \cdot \qqv \right) + \ppder{s}{\nabla \qrho} \cdot \nabla \qqv \right] + \Tensor{\mathcal{J}},
\end{align}
which (in general, just like $ \qqP $, $ \qqJE $ and $ f $) depends on all elements of the constitutive state space, however, the so-called residual entropy current density $ \Tensor{\mathcal{J}} $---it is usually assumed to be zero---depend only on the variables $ \left( \qrho, \nabla \qrho, \qqv, e, \qphi, \nabla \qphi \right) $. Replacing the Lagrange--Farkas multipliers obtained from the Liu-equations \re{eq:Liu-1}--\re{eq:Liu-8} and the entropy current density \re{eq:JS-nonloc} into \re{eq-s-ineq-nonloc}, after rearranging one obtains
\begin{align}
	\nonumber
	0 &\le \pd_j \left( \pder{s}{\qv^i} \right) \qP^{ij} + \pd_j \left( \pder{s}{e} \right) \qJE^j - \left[ \qrho \pder{s}{\qphi} - \pd_j \left( \qrho \ppder{s}{\pd_j \qphi} \right) \right] f \\
	\label{eq:dis-ineq-nonloc-ind}
	& \hskip 2.7ex
	- \left[ \qrho^2 \pder{s}{\qrho} \delta_i^j - \frac{\qrho^2}{2} \pd_k \left( \ppder{s}{\pd_l \qrho} \right) \left( \delta_l^k \delta_i^j + \delta_l^j \delta_i^k \right) + \qrho \ppder{s}{\pd_j \qphi} \pd_i \qphi \right] \pd_j \qv^i,
\end{align}
\ie in the invariant notations
\begin{align}
	\nonumber
	0 &\le \nabla \left( \pder{s}{\qqv} \right) : \qqP + \nabla \left( \pder{s}{e} \right) \cdot \qqJE - \left( \qrho \pder{s}{\qphi} - \nabla \cdot \left( \qrho \ppder{s}{\nabla \qphi} \right) \right) f \\
	\label{eq:dis-ineq-nonloc}
	& \hskip 2.7ex
	- \left[ \qrho^2 \left( \pder{s}{\qrho} - \frac{1}{2} \nabla \cdot \left( \ppder{s}{\nabla \qrho} \right) \right) \ident - \frac{\qrho^2}{2} \nabla \left( \ppder{s}{\nabla \qrho} \right) + \qrho \ppder{s}{\nabla \qphi} \otimes \nabla \qphi \right] : \nabla \qqv,
\end{align}
where, throughout the paper, tensor products are written according to the index convention $ \left( \ttensor{a} \otimes \ttensor{b} \right)_{ji} = a_j b_i $. Here we add two important remarks:
\begin{itemize}
	\item Entropy production rate density \re{eq:dis-ineq-nonloc} consists of the sum of four terms, however, only three constitutive functions appear in it. The separation of internal and kinetic energies will reduce these four terms to three independent, purely constitutional terms, which will be detailed in the next subsection.
	\item Even though there were no variational principles or formulation applied, a partial functional derivative of the entropy appears in \re{eq:dis-ineq-nonloc}. Since $ \qrho $ and $ \qphi $ are independent variables, the third term in \re{eq:dis-ineq-nonloc} can be rearranged as
	\begin{align}
		\label{eq:dis-ineq-nonloc-invar}
		\qrho \pder{s}{\qphi} - \nabla \cdot \left( \qrho \ppder{s}{\nabla \qphi} \right) = \pder{\left( \qrho s \right)}{\qphi} - \nabla \cdot \left( \ppder{\left( \qrho s \right)}{\nabla \qphi} \right) = \frac{\delta}{\delta \qphi} \int\limits_{\mathcal{V}} \qrho s \left( \qrho, \nabla \qrho, \qqv, e, \qphi, \nabla \qphi \right) \dd V = \frac{\delta S}{\delta \qphi}.
	\end{align}
	Such partial functional derivatives related to weakly nonlocal variables will continue to appear.
\end{itemize}

\subsection{Energy representation}

According to objectivity arguments, material properties cannot depend on relative velocity of the material and the laboratory frame \cite{TruNol65b,CimEta14a}. This observation is reflected in the fact that entropy, as a constitutive function depends only on internal energy, not on total energy. However, this simple argument does not consider the possible frame dependence of the other thermodynamic state variables, it must be extended in a proper frame independent treatment. A justification leads to a covariant, frame independent formulation of the evolution equations and the material properties as well. In the Galilean relativistic spacetime model one can recover the internal energy as a particular four-tensor component comoving with the material \cite{Van17a}. Here, in the following we restrict ourself to the traditional idea and concept of internal energy.

Via the relationship between the total and internal energy \re{eq:tot-int-kin}, we may write that $ s \left( \qrho, \nabla \qrho, \qqv, e, \qphi, \nabla \qphi \right) = \tilde{s} \big( \qrho, \nabla \qrho, u \left( e, \qqv \right), \qphi, \nabla \qphi \big)$, hence
\begin{align}
	\pder{s}{e} &= \pder{\tilde{s}}{u}, & 
	\pder{s}{\qqv} &= - \pder{\tilde{s}}{u} \qqv.
\end{align}
Therefore, \re{eq:dis-ineq-nonloc} can be rewritten as (omitting tildes for clarity) 
\begin{align}
	\nonumber
	0 &\le \nabla \left( \pder{{s}}{u} \right) \cdot (\qqJE -\qqv\cdot \qqP)- \left( \qrho \pder{s}{\qphi} - \nabla \cdot \left( \qrho \ppder{s}{\nabla \qphi} \right) \right) f \\
	\label{eq:dis-ineq-nonloc-solv}
	& \hskip 2.7ex
	- \left( \pder{{s}}{u}\qqP + \qrho^2 \left( \pder{s}{\qrho} - \frac{1}{2} \nabla \cdot \left( \ppder{s}{\nabla \qrho} \right) \right) \ident - \frac{\qrho^2}{2} \nabla \left( \ppder{s}{\nabla \qrho} \right) + \qrho \ppder{s}{\nabla \qphi} \otimes \nabla \qphi \right) : \nabla \qqv.
\end{align}

That is, one obtains a quadratic expression with three terms: thermal, internal variable and mechanical ones [which are represented consecutively in \re{eq:dis-ineq-nonloc-solv}] with constitutive functions in each terms, namely, $\qqJE$, $f$ and $\qqP$, respectively. Also one can recognise the \emph{internal energy current density} $\qqJU=\qqJE -\qqv\cdot \qqP$ in the thermal term.

According to classical thermodynamics, the specific entropy is a function of the specific \emph{fluid internal energy} $ u \qf $ and the specific volume $v = \frac{1}{\qrho}$, \ie $ s = s \left( u \qf, \qrho \right)$. The partial derivatives of specific entropy define the absolute temperature $ T $ and the hydrostatic fluid pressure $ p \qf $, which is expressed through the Gibbs relation [\cf \re{eq:Gibbs-s}]
\begin{align}
	\label{eq:s-Gibbs}
	\dd s = \frac{1}{T} \dd u \qf - \frac{1}{\qrho^2} \frac{p \qf}{T} \dd \qrho,
\end{align}
furthermore, extensivity implies the specific Euler relation [\cf \re{eq:S-pder-2}]
\begin{align}
	\label{eq:s-Euler}
	s = \frac{1}{T} u \qf + \frac{p \qf}{T} \frac{1}{\qrho} - \frac{\mu \qf}{T}
\end{align}
with the fluid chemical potential $ \mu \qf $. Accordingly, the constitutive relation for the mechanical term (characterized by $ p \qf $) in the entropy representation is not purely mechanical, due to the multiplication by the reciprocal of the temperature. However, if one transforms the Gibbs relation \re{eq:s-Gibbs} into energy representation, \ie specific fluid internal energy is assumed to be the function of specific entropy and density, then the differential of specific fluid internal energy can be written as
\begin{align}
	\label{eq:u-Gibbs}
	\dd u \qf = T \dd s + \frac{1}{\qrho^2} p \dd \qrho,
\end{align}
thus the two terms on the r.h.s. of \re{eq:u-Gibbs} purely characterize the thermal and mechanical interactions.

Further interactions can be described via new variables (\eg internal variables), hence the thermodynamic state space should be extended and, all these new variables may have a corresponding internal energy contribution. In parallel, any gradient extension of existing interactions may also result in an internal energy contribution. All these internal energy contributions can be separated from the internal energy and what remains is the fluid internal energy. Therefore it is advisable to use the following separated form: 
\begin{align}
	\label{eq:e-nonloc}
	u \qf =
	u - \frac{\he \left( \qrho, \nabla \qrho, \qphi, \nabla \qphi \right) }{\qrho},
\end{align}
where $ \he $ is the energy density contribution of the internal variable and the gradients. In the following will refer $\he$ as \emph{\adden \ density}. 

Therefore, specific entropy can be expressed by the various variables as $ s \left( u, \qrho, \nabla \qrho, \qphi, \nabla \qphi \right) = s \big( u \qf \left( u, \qrho, \nabla \qrho, \qphi, \nabla \qphi \right), \qrho \big) = s \big( u - \frac{\he \left( \qrho, \nabla \qrho, \qphi, \nabla \qphi \right)}{\qrho}, \qrho \big) $ and then the Gibbs relation reads as
\begin{align}
	\label{eq:Gibbs-nonloc}
	\dd s = \frac{1}{T} \dd u - \frac{1}{T\qrho} \left( \frac{p \qf-\he}{\qrho} + \pder{\he}{\qrho} \right) \dd \qrho - 
	\frac{1}{T\qrho} \ppder{\he}{\nabla \qrho} \cdot \dd \left( \nabla \qrho \right) - 
	\frac{1}{T\qrho} \pder{\he}{\qphi} \dd \qphi - 
	\frac{1}{T\qrho} \ppder{\he}{\nabla \qphi} \cdot \dd \left( \nabla \qphi \right),
\end{align}
from which the partial derivatives of specific entropy and specific internal energy are
\begin{align}
	\label{eq:pder-1}
	\pderr{s}{u}{\qrho, \nabla \qrho, \qphi, \nabla \qphi } &= \frac{1}{T}, &
	\pderr{u}{s}{\qrho, \nabla \qrho, \qphi, \nabla \qphi } &= T, \\
	\pderr{s}{\qrho}{u, \nabla \qrho, \qphi, \nabla \qphi } &= - \frac{1}{T\qrho} \left( \frac{p \qf -\he}{\qrho} + \pder{\he}{\qrho} \right), &
	\pderr{u}{\qrho}{s, \nabla \qrho, \qphi, \nabla \qphi } &= \frac{1}{\qrho} \left( \frac{p \qf -\he}{\qrho} + \pder{\he}{\qrho} \right),\\
	\ppderr{s}{\nabla \qrho}{u, \qrho, \qphi, \nabla \qphi } &= - \frac{1}{T\qrho} \ppder{\he}{\nabla \qrho}, &
	\ppderr{u}{\nabla \qrho}{s, \qrho, \qphi, \nabla \qphi } &= \frac{1}{\qrho}\ppder{\he}{\nabla \qrho}, \\
	\pderr{s}{\qphi}{u, \qrho, \nabla \qrho, \nabla \qphi } &= - \frac{1}{T\qrho} \pder{\he}{\qphi}, &
	\pderr{u}{\qphi}{s, \qrho, \nabla \qrho, \nabla \qphi } &= \frac{1}{\qrho}\pder{\he}{\qphi} \\
	\label{eq:pder-5}
	\ppderr{s}{\nabla \qphi}{u, \qrho, \nabla \qrho, \qphi } &= - \frac{1}{T\qrho} \ppder{\he}{\nabla \qphi}, &
	\ppderr{u}{\nabla \qphi}{s, \qrho, \nabla \qrho, \qphi } &= \frac{1}{\qrho} \ppder{\he}{\nabla \qphi}.
\end{align}
Via the partial derivatives \re{eq:pder-1}--\re{eq:pder-5} the dissipation inequality \re{eq:dis-ineq-nonloc-solv} can be transformed into energy representation, hence one obtains
\begin{align}
	\nonumber
	0 &\le
	\frac{1}{T} \left( \pder{\he}{\qphi} - \nabla \cdot \left( \ppder{\he}{\nabla \qphi} \right) \right) f +\left( \qqJE - \qqP \cdot \qqv - \frac{\qrho}{2} \left( \ppder{\he}{\nabla \qrho} \left( \nabla \cdot \qqv \right) + \ppder{\he}{\nabla \qrho} \cdot \nabla \qqv \right) - \ppder{\he}{\nabla \qphi} f \right) \cdot \nabla \frac{1}{T} \\
	\label{eq:dis-ineq-nonloc-1}
	& \hskip 2.7ex
	- \frac{1}{T} \left( \qqP - \left( p \qf - \he + \qrho \pder{\he}{\qrho} - \frac{\qrho^2}{2} \nabla \cdot \left( \frac{1}{\qrho} \ppder{\he}{\nabla \qrho} \right) \right) \ident + \frac{\qrho^2}{2} \nabla \left( \frac{1}{\qrho} \ppder{\he}{\nabla \qrho} \right) - \ppder{\he}{\nabla \qphi} \otimes \nabla \qphi \right) : \nabla \qqv.
\end{align}
Coefficients of (inverse) temperature gradient and velocity gradient define heat current density or heat flux
\begin{align}
	\label{eq:q}
	\qqq = \qqJE - \qqP \cdot \qqv - \frac{\qrho}{2} \left( \ppder{\he}{\nabla \qrho} \left( \nabla \cdot \qqv \right) + \ppder{\he}{\nabla \qrho} \cdot \nabla \qqv \right) - \ppder{\he}{\nabla \qphi} f
\end{align}
and viscous pressure tensor
\begin{align}
	\label{eq:Pi}
	\qqPi = \qqP - \left( p \qf -\he + \qrho\pder{\he}{\qrho} - 
	\frac{\qrho}{2} \nabla \cdot \left( \pder{\he}{\nabla \qrho} \right) + 
	\frac{1}{2} \pder{\he}{\nabla \qrho} \cdot \nabla \qrho\right) \ident + 
	\frac{\qrho}{2} \nabla \left( \pder{\he}{\nabla \qrho} \right) - 
	\frac{1}{2}\pder{\he}{\nabla \qrho} \otimes \nabla \qrho-
	\pder{\he}{\nabla \qphi} \otimes \nabla \qphi, 
\end{align}
respectively. Then, replacing \re{eq:q} and \re{eq:Pi} into \re{eq:JS-nonloc}{, and adopting the conventional choice of vanishing residual entropy current, $ \Tensor{\mathcal{J}} = \zero $,} one finds the well-known entropy flux--heat flux relationship expression as 
\begin{align}
	\qqJS = \frac{\qqq}{T}.
\end{align}
One can see the role of dissipation, determined by the entropy balance, and the role of energy dispersion determined by the internal energy balance. The latter one is obtained by constraining the balance of total energy \re{eq:bal-e} with the balance of linear momentum \re{eq:bal-v}, \ie
\begin{align}
	\label{eq:bal-u}
	\qrho \dot u + \nabla \cdot \left[ \qqq + \frac{\qrho}{2} \left( \ppder{\he}{\nabla \qrho} \left( \nabla \cdot \qqv \right) + \ppder{\he}{\nabla \qrho} \cdot \nabla \qqv \right) + \ppder{\he}{\nabla \qphi} f \right] = - \qqP : \nabla \qqv,
\end{align}
hence, conductive current density of internal energy $\qqJU$ is not identified with the heat current density $\qqq$, \ie
\begin{align}
	\label{eq:JU}
	\qqJU = \qqJE - \qqP \cdot \qqv = \qqq + \frac{\qrho}{2} \left( \ppder{\he}{\nabla \qrho} \left( \nabla \cdot \qqv \right) + \ppder{\he}{\nabla \qrho} \cdot \nabla \qqv \right) + \ppder{\he}{\nabla \qphi} f,
\end{align}
and the source terms is the power of mechanical work originated from the complete local pressure. Therefore, the usual form of internal energy 
\begin{align}
	\label{eq:bal-u-int}
	\qrho \dot u + \nabla \cdot \qqJU = - \qqP : \nabla \qqv 
\end{align}
is obtained.

Let us give here a short technical remark. To determine entropy production rate density via the method of divergence separation---\ie via the heuristic method of Classical Irreversible Thermodynamics---, one assumes a specific form on the specific entropy, in our investigated problem it is considered to be $ s = s \left(u - \frac{\he \left( \qrho, \nabla \qrho, \qphi, \nabla \qphi \right)}{\qrho}, \qrho\right) $. Then the substantial time derivative of this specific entropy function is determined, which is directly constrained by the balance equations of mass and internal energy. However, usually the balance of internal energy is written as
\begin{align}
	\label{eq:bal-u-ass}
	\qrho \dot u + \nabla \cdot \qqq = - \qqP : \nabla \qqv,
\end{align}
which results in a generalized entropy current density -- heat current density relationship in the form of $ \qqJS = \frac{1}{T} \left( \qqq - \hat \qqq \right) $ with an additional energy current density $ \hat \qqq $. Introducing the concept of interstitial working, the balance of internal energy may be given as
\begin{align}
	\label{eq:bal-u-ass-DS}
	\qrho \dot u + \nabla \cdot \left( \qqq + \hat \qqq \right) = - \qqP : \nabla \qqv,
\end{align}
hence a modified energy current density is assumed. Repeating the calculations via \re{eq:bal-u-ass-DS} the classical $ \qqJS = \frac{1}{T} \qqq $ entropy flux---heat flux relationship is obtained. This means that the previously mentioned heuristic methods do not provide sufficient insight. Although in the light of our previous, more rigorous derivation, $ \hat \qqq $ can be identified, the heuristic methods do not reveal that $ \qqJU $ is actually the internal energy current density, which can not be directly identified with the heat current density in such problems when not just the fluid internal energy forms the total internal energy. 

\subsection{Linear constitutive equations}

{As already demonstrated for the entropy functional in
\re{eq:dis-ineq-nonloc-invar}, partial functional derivatives naturally arise from weakly nonlocal dependencies. Accordingly, in the energy representation, we introduce the notation $ \frac{\delta \he}{\delta \qphi} = \pder{\he}{\qphi} - \nabla \cdot \left( \ppder{\he}{\nabla \qphi} \right) $ for a more transparent presentation of the formulas.} Applying \re{eq:q} and \re {eq:Pi} the entropy production density rate \re{eq:dis-ineq-nonloc-1} is written as 
\begin{align}
	\label{eq:dis-ineq-nonloc-2}
	0 &\le \frac{1}{T} \frac{\delta \he}{\delta \qphi} f + \qqq \cdot\nabla\frac{1}{T} - \qqPi : \frac{\nabla \qqv}{T} \\
	\nonumber
	&= \frac{1}{T} \frac{\delta \he}{\delta \qphi} f + \qqq \cdot \nabla \frac{1}{T} - \frac{1}{3 T} \pi\left( \nabla \cdot \qqv \right) - \frac{1}{T} \langle \qqPi \rangle : \langle \nabla \qqv \rangle - \frac{1}{T} \left( \qqPi \right)^{\rm Skw} : \left( \nabla \qqv \right)^{\rm Skw},
\end{align}
where the isotropic representation of the viscous pressure tensor, \ie
\begin{align}
	\qqPi &= \qqPi ^{\rm Sym} + \qqPi ^{\rm Skw} = \frac{1}{3} \pi \ident + \langle \qqPi \rangle + \qqPi ^{\rm Skw}, && \text{where} &
	\pi &= \operatorname{tr} \qqPi 
\end{align}
is applied, with $ \operatorname{tr} $ denoting the trace of a second order tensor. $ \ttensor{A}^{\rm Sym} $, $ \langle \ttensor{A} \rangle $ and $ \ttensor{A}^{\rm Skw} $ are the symmetric, the symmetric traceless, \ie the deviatoric and, the skew-symmetric parts of the second order tensor $ \ttensor{A} $, respectively. 

Following the idea of Onsager, positive semi-definiteness of \re{eq:dis-ineq-nonloc-2} is ensured via linear equations, which is the simplest solution of the inequality. Assuming an isotropic fluid, the different tensorial orders and characters do not couple (which is known as Curie's principle), hence the linear Onsagerian equations can be prescribed on the isotropic independent parts of the constitutive functions---usually called \emph{thermodynamic fluxes}---, which are treated as linear functions of the so-called \emph{thermodynamic forces}. This kind of distinction is presented in Tab.~\ref{tab:1}.
\begin{table}[h!]
	\centering
	\begin{tabular}{c | c | c | c }
		Interaction type & Thermodynamic flux & Thermodynamic force & Tensorial character \\
		\hline
		Internal variable & $ f $ & $ \frac{\delta \he}{\delta \qphi} $ & scalar \\
		Thermal & $ \qqq $ & $ \nabla \frac{1}{T} $ & vector \\
		Mechanical & $ \pi $ & $ \left( - \nabla \cdot \qqv \right) $ & scalar \\
		Mechanical & $ \langle \qqPi \rangle $ & $ \langle - \nabla \qqv \rangle $ & (deviatoric) tensor \\
		Mechanical & $ \left( \qqPi \right)^{\rm Skw} $ & $ \left( - \nabla \qqv \right)^{\rm Skw} $ & axial vector
	\end{tabular}
	\caption{Thermodynamic fluxes and forces.}
	\label{tab:1}
\end{table}

Although skew-symmetric second-order tensors in the 3-dimensional space can be represented as axial vectors, coupling between vectors and axial vectors is not possible in isotropic materials since axial vectors actually are second-order tensors, hence they transform differently than vectors. Therefore, in our case, only the two scalar quantities can be coupled. Based on the previous statements the linear Onsagerian equations, \ie the linear flux-force relations are
\begin{align}
	\label{eq:Ons-1}
	\frac{1}{3} \pi &= \eta_{\rm Vol} \left( - \nabla \cdot \qqv \right) + L_{\pi f} \frac{\delta \he}{\delta \qphi}, \\
	\label{eq:Ons-2}
	f &= L_{f \pi } \left( - \nabla \cdot \qqv \right) + L_{ff} \frac{\delta \he}{\delta \qphi}, \\
	\label{eq:Ons-3}
	\qqq &= \lambda \nabla \frac{1}{T}, \\
	\label{eq:Ons-4}
	\langle \qqPi \rangle &= 2 \eta \langle - \nabla \qqv \rangle, \\
	\label{eq:Ons-5}
	\left( \qqPi \right)^{\rm Skw} &= 2 \eta_{\rm Rot} \left( - \nabla \qqv \right)^{\rm Skw} 
\end{align}
with volumetric, shear and rotational viscosities $ \eta_{\rm Vol} $, $ \eta $ and $ \eta_{\rm Rot} $, thermal conductivity $ \lambda $ and coefficients $ L_{ \pi f } $, $ L_{ f \pi } $ and $ L_{ f f } $. According to \re{eq:Ons-1}--\re{eq:Ons-5} the dissipation inequality \re{eq:dis-ineq-nonloc-2} can be given in the quadratic form
\begin{align}
	\label{eq:ent-pr-mat}
	0 \le \frac{1}{T} \begin{pmatrix} \left( -\nabla \cdot \qqv \right) & \frac{\delta \he }{\delta \qphi} & T \nabla \frac{1}{T} & \langle - \nabla \qqv \rangle & \left( - \nabla \qqv \right)^{\rm Skw} \end{pmatrix}
	\begin{pmatrix} \eta_{\rm Vol} & \frac{L_{\pi f} + L_{f \pi}}{2} & 0 & 0 & 0\\
		\frac{L_{\pi f} + L_{f \pi}}{2} & L_{ff} & 0 & 0 & 0 \\
		0 & 0 & \lambda & 0 & 0 \\ 0 & 0 & 0 & 2 \eta & 0 \\ 0 & 0 & 0 & 0 & 2 \eta_{\rm Rot} \end{pmatrix}
	\begin{pmatrix} \left( - \nabla \cdot \qqv \right) \\ \frac{\delta \he }{\delta \qphi} \\ \nabla \frac{1}{T} \\ \langle - \nabla \qqv \rangle \\ \left( - \nabla \qqv \right)^{\rm Skw} \end{pmatrix}.
\end{align}
Therefore, by Sylvester's criteria the conditions
\begin{align}
	\lambda & \ge 0, &
	\eta & \ge 0, &
	\eta_{\rm Vol} & \ge 0, &
	\eta_{\rm Rot} & \ge 0, &
	L_{ f f } & \ge 0, &
	\eta_{\rm Vol} L_{ f f } - \left( \frac{L_{\pi f} + L_{f \pi}}{2} \right)^2 &\ge 0
	\label{eq:ineq}\end{align}
follows. In general, the above defined transport coefficients can depend on all elements of the constitutive state space. It is a simple consequence of Lagrange's mean value theorem. Since we rely on the case of a symmetric pressure tensor, the antisymmetric part of its viscous contribution must vanish, \ie $ \eta_{\rm Rot} = 0 $. Even so, let us remark here that the antisymmetric part of the pressure tensor plays a role in polar fluids, where the internal angular momentum of the fluid is not negligible \cite{TruTou60b}. Polar fluids appeared as models of liquid crystals \cite{Ver97b}, and recently, the observed spin polarization of quark gluon plasma has increased interest in relativistic spin fluids, \eg \cite{BecTin10a,FloEta18a}. 

Without deeper investigations there is no reason to assume any kind of reciprocal relations, \ie the relationship between $L_{\pi f}$ and $L_{f\pi}$. Reciprocal relations may be justified with microscopic or macroscopic time-reversal symmetry \cite{PavEta14a}, when the type (\ie $ \alpha $ or $ \beta $) of the variables are known. However, in general, the validity of reciprocal relations is questionable in the case of internal variables, where the thermodynamic compatibility of the evolution equations is the only requirement.

\subsection{Balance equation of fluid internal energy}

Finally, in light of the determined constitutive functions [see equations \re{eq:Ons-1}--\re{eq:Ons-5}] let us present the balance equation of fluid internal energy. We remind the reader that the derivation considered non-polar fluids, \ie pressure tensor is symmetric.

Internal energy is expressed via \re{eq:e-nonloc}, therefore, balance of internal energy \re{eq:bal-u} is constrained by the balance of mass \re{eq:bal-m}, the evolution equation of the internal variable \re{eq:xi-evol} and the evolution equations of gradients of density and internal variable, \re{eq:gradrho-evol} and \re{eq:gradxi-evol}, thus, the balance of fluid internal energy becomes 
\begin{align}
	\qrho \dot u \qf + \nabla \cdot \qqq = - \left( p \qf \ident + \qqPi \right) : \nabla \qqv + \left[ \pder{\he}{\qphi} - \nabla \cdot \left( \ppder{\he}{\nabla \qphi} \right) \right] f.
\end{align}
Applying the Onsagerian equations \re{eq:Ons-1}, \re{eq:Ons-2} and \re{eq:Ons-4} we find
\begin{align}
	\label{eq:bal-u-fl}
	\qrho \dot u \qf = - \nabla \cdot \qqq - p \qf \left( \nabla \cdot \qqv \right) + \underbrace{ \left( \eta_{\rm Vol} \left( \nabla \cdot \qqv \right)^2 + 2 \eta \langle \nabla \qqv \rangle : \langle \nabla \qqv \rangle \right) \vphantom{\left( \frac{\delta \left( \qrho \he \right)}{\delta \qphi} \right)^2}}_{\text{viscous dissipation}} \underbrace{ - \left( L_{\pi f} + L_{f \pi } \right) \left( \nabla \cdot \qqv \right) \frac{\delta \he }{\delta \qphi} + L_{ff} \left( \frac{\delta \he }{\delta \qphi} \right)^2}_{\text{additional volumetric source}}.
\end{align}
Comparing equation \re{eq:bal-u-fl} to \re{eq:bal-u-ass}---which is actually the balance of internal energy for pure fluids without extra state variables---, additional source terms---caused by the internal variable---appear. Furthermore, the conductive current of fluid internal energy can be identified via heat current density.

\section{Perfect fluids} \label{sec:per-flu}

In perfect fluids, there is no dissipation, therefore, entropy production rate density is zero. When we restrict ourselves to a cross-effect free treatment, \ie the coefficients $L_{\pi f}$ and $L_{f\pi}$ in equations \re{eq:Ons-1} and \re{eq:Ons-2} are assumed to be zero, this means that the previously defined constitutive functions, \ie $ f $, $ \qqq $, and $ \qqPi $ must be zero. This highlights that one may distinguish various levels of perfectness according to the different terms in the entropy production \re{eq:dis-ineq-nonloc-2}.

In an \textit{internally} perfect fluid $ f = 0 $, hence the scalar field $\varphi$ does not change in a local rest frame. Therefore 
\begin{align}
	\label{eq:GL}
	0 = \pder{\he}{\qphi} - \nabla \cdot \left( \ppder{\he}{\nabla \qphi} \right).
\end{align}

A \textit{mechanically} perfect fluid is characterised by zero mechanical dissipation, therefore, the perfect fluid's pressure tensor is
\begin{align}
	\label{eq:P_id}
	\qqP \id = \left( p \qf -\he + \qrho \pder{\he}{\qrho} - 
	\frac{\qrho^2}{2} \nabla \cdot \left( \frac{1}{\qrho}\ppder{\he}{\nabla \qrho} \right)\right) \ident - \frac{\qrho^2}{2} \nabla \left( \frac{1}{\qrho}\ppder{\he}{\nabla \qrho} \right) +
	\ppder{\he}{\nabla \qphi} \otimes \nabla \qphi.
\end{align}
Note that the gradients of the density and the scalar field also contribute to the pressure tensor through the extra energy density $ \he $, which is resulted in a much richer behavior than for a classical fluid.
{
Although the constitutive expression for the perfect-fluid pressure tensor \re{eq:P_id} follows from the entropy inequality, pressure symmetry is an additional requirement associated with the non-polar character of the fluid. Therefore, in the following, we restrict ourself to the extra energy densities compatible with $\qqP\id=(\qqP\id)^{\rm T}$. We do not derive the resulting condition in its most general form, but enforce it explicitly for the constitutive specializations considered in Sec.~\ref{sec:spec}.
}

Finally, in a \textit{thermally} perfect fluid the heat current density $\qqq$ is zero, which---based on \re{eq:JU}---is resulted in the nonzero internal energy current density 
\begin{align}
	\label{eq:JU-therm-per}
	\qqJU = \frac{\qrho}{2} \left( \ppder{\he}{\nabla \qrho} \left( \nabla \cdot \qqv \right) + \ppder{\he}{\nabla \qrho} \cdot \nabla \qqv \right) + \ppder{\he}{\nabla \qphi} f.
\end{align}
Note that the right-hand side depends on the velocity field, as well as on the internal variable. Therefore, in contrast with internally or mechanically perfect fluids, which are characterized by static equilibrium conditions (that are determined purely by the entropy function, or in our case the \adden\ density function $\he$), thermal perfectness is a dynamical property as it depends on the velocity field. When a fluid is both internally perfect and thermally perfect, then in the light of \re{eq:GL} the internal energy current density \re{eq:JU-therm-per} reduces to
\begin{align}
	\qqJU = \frac{\qrho}{2} \left( \ppder{\he}{\nabla \qrho} \left( \nabla \cdot \qqv \right) + \ppder{\he}{\nabla \qrho} \cdot \nabla \qqv \right).
\end{align}

It is remarkable that the fluid can also be perfect if Casimir reciprocity is assumed for the cross-effects, \ie when $ L_{\pi f} = - L_{f \pi} $, since entropy production rate density remains zero in this case, \cf \re{eq:ent-pr-mat}. Therefore, an additional term originated from the internal variable emerges in the pressure tensor and, in parallel, a scalar pressure---through the divergence of the velocity field---contributes to the field equation of the internal variable. Although this coupling may result in particular physical consequences (see \eg in \cite{VanAbe22a,AbeVan22a}), we do not further investigate this case, it is left to a future work.

\subsection{Emerging conservative volumetric force field}

An interesting observation related to perfect fluids is that the balance of linear momentum \re{eq:bal-v} can be reformulated into an Euler equation with a conservative volumetric force field. First, let us calculate the divergence of the pressure tensor, that is
\begin{align}
	\nonumber
	\partial_j \left( \qP \id \right)_i^j &= \partial_j p \qf + \partial_j\left( \left(-\he + \qrho \pder{\he}{\qrho} - 
	\frac{\qrho^2}{2} \partial_k \left( \frac{1}{\qrho}\ppder{\he}{\partial_k \qrho} \right)\right)\delta^{ij} - 
	\frac{\qrho^2}{2} \partial_i \left( \frac{1}{\qrho}\ppder{\he}{\partial_j \qrho} \right) +
	\ppder{\he}{\partial_j \qphi} \partial_i\qphi \right) \\
	\nonumber 
	&= \partial_j p \qf +
	\underbrace{\qrho \partial_i \pder{\he}{\qrho} - \ppder{\he}{\partial_k\qrho}\partial_{ik}\qrho - \pder{\he}{\qphi}\partial_{i}\qphi - \ppder{\he}{\partial_k\qphi}\partial_{ik}\qphi}_{ = \partial_i\left(-\he + \qrho \pder{\he}{\qrho}\right)} \\
	\nonumber
	&\hskip 2.7ex
	\underbrace{ - \frac{1}{2} \pd_k \ppder{\he}{\pd_k \qrho} \pd_i \qrho - \frac{\qrho}{2} \pd_{ik} \ppder{\he}{\partial_k \qrho} + \frac{1}{2} \pd_i \ppder{\he}{\partial_k \qrho} \pd_k \qrho + \frac{1}{2} \ppder{\he}{\partial_k \qrho} \pd_{ik} \qrho }_{=\pd_i \left( - 
		\frac{\qrho^2}{2} \partial_k \left( \frac{1}{\qrho}\ppder{\he}{\partial_k \qrho} \right) \right)} \\
	\nonumber
	&\hskip 2.7ex
	\underbrace{ - \frac{1}{2} \pd_i \ppder{\he}{\pd_j \qrho} \pd_j \qrho - \frac{\qrho}{2} \pd_{ij} \ppder{\he}{\partial_j \qrho} + \frac{1}{2} \pd_j \ppder{\he}{\partial_j \qrho} \pd_i \qrho + \frac{1}{2} \ppder{\he}{\partial_j \qrho} \pd_{ij} \qrho }_{=\pd_j \left( - 
		\frac{\qrho^2}{2} \partial_i \left( \frac{1}{\qrho}\ppder{\he}{\partial_j \qrho} \right) \right)}\\
	\nonumber
	& \hskip 2.7ex 
	\underbrace{+ \pd_j \ppder{\he}{\pd_j \qphi} \pd_i \qphi + \ppder{\he}{\pd_j \qphi} \pd_{ij} \qphi}_{=
		\partial_j\left(\ppder{\he}{\partial_j \qphi}\partial_i \qphi \right)}\\ 
	&=	\partial_j p \qf +
	\qrho \partial_i \left(\pder{\he}{\qrho} - \partial_j \ppder{\he}{\partial_j \qrho}\right)- 
	\partial_i\qphi\left(\pder{\he}{\qphi} -\partial_j\ppder{\he}{\partial_j \qphi} \right) =	\partial_j p \qf + \qrho \partial_i \frac{\delta \he}{\delta \qrho} - \partial_i\qphi\frac{\delta \he}{\delta \qphi}.
\end{align} 
Thereby, the divergence of the pressure tensor in invariant notation is 
\begin{align}
	\label{eq:holop}
	\nabla \cdot \qqP \id = \nabla p \qf + \qrho \nabla \frac{\delta \he}{\delta \qrho} - \frac{\delta \he}{\delta \qphi} \nabla \qphi.
\end{align}
If the fluid is internally perfect, \eg the internal variable relaxes to equilibrium faster than the hydrodynamic degrees of freedom, then, according to \re{eq:GL}, \re{eq:holop} reduces to
\begin{align}
	\label{eq:holop-1}
	\nabla \cdot \qqP \id = \nabla p \qf + \qrho \nabla \frac{\delta \he}{\delta \qrho}.
\end{align}
Substituting \re{eq:holop-1} into the balance of linear momentum \re{eq:bal-v} the Euler equation
\begin{align}   
	\label{eq:euler}
	\qrho \dot \qqv &= - \nabla p \qf + \qrho \ttensor{f}_{\rm vol},
\end{align}
is obtained, with the conservative volumetric force field $ \ttensor{f}_{\rm vol} = - \nabla \frac{\delta \he}{\delta \qrho} $. We emphasize that originally, at the beginning of the derivation, we did not take into account any volumetric source term, it emerges in \re{eq:euler} as a direct mechanical manifestation of the additional variables encoded in the extra energy density. Importance of this recognition is further analyzed in Sec.~\ref{subsec:ng}. In the reverse case, \ie when transforming a conservative force field into the pressure tensor, seems much more natural, for instance, when gravitational potential is interpreted as a pressure contribution in case of buoyancy flows. However, as we will present it in Sec.~\ref{subsec:kor}, in case of more general potentials containing also weakly nonlocal terms, the scalar potential can be interpreted with different pressure tensors, which are equivalent up to a full divergence term.

{Let us finally comment on the role of internal perfectness. If the equilibrium condition is not satisfied, \ie $ \frac{\delta\he}{\delta\qphi} \neq 0 $, the last term in \re{eq:holop} remains as an additional volumetric force associated with the nonequilibrium state of $ \qphi $. This distinction is essential for the holographic reduction: only part of the interaction can then be represented by the density-weighted gradient of a scalar potential, whereas the remainder term is an explicit bulk-force contribution.
}

\subsection{Classical holography}
\label{sec:clahol}
Now, we are ready to formulate the conditions of the holographic property of ideal continua (see \re{qholodem}), considering thermodynamic constraints from the Second Law of Thermodynamics. A field theory is called \emph{classical holographic}, if there is a scalar potential $\Phi$ related to the perfect fluid pressure so that 
\begin{align}
	\label{eq:cl-holo}
	\nabla \cdot \qqP \id &= \qrho \nabla \Phi
\end{align}
identity holds. 

{Since the terminology invites comparison with holography in the field-theoretic and gravitational sense, we state at once what is and what is not intended. Throughout this paper, \emph{classical holography} means the scalar-potential representation \re{eq:cl-holo} of $\nabla\cdot\qqP\id$ and nothing stronger. The holographic principle of 't~Hooft and Susskind~\cite{tHo93a,Sus95a} is a statement about the \emph{number} of degrees of freedom, namely that the information content of a bulk region is bounded by its boundary area at roughly one bit per Planck area; in the AdS/CFT correspondence it is realized as a duality between a gravitational bulk and a lower-dimensional quantum field theory, of which the fluid--gravity correspondence~\cite{BhaEta08a} is a long-wavelength limit. The property considered here is categorically different and much more modest: an exact local identity of non-relativistic continuum thermodynamics, by virtue of which the bulk force density can be expressed, through constitutive properties of perfect continua, as a traction on the boundary of any region. 

The kinship that justifies the name is nevertheless substantive, and it was identified in a related context by Hossenfelder~\cite{Hos10a}, who showed that the classical and the thermodynamic-holographic formulations of static Newtonian gravity are equivalent, the equivalence resting on the integral identities of Gauss and Green for harmonic functions. What is holographic about static Newtonian gravity, on that reading, is a property of harmonic functions rather than of Planck-scale physics, and theories of this kind are better described as holography-friendly than as holographic. The present result may be read as the continuum-thermodynamic counterpart of that observation and as an extension of it in two respects: the boundary-encodes-bulk structure is extended beyond pure gravity, as shown below, and it follows from the Second Law rather than from the harmonicity of a particular field. We note in this connection that \cite{CarRem16a}, distinguishing holographic from thermodynamic routes to emergent gravity, identify as an obstacle to the latter the difficulty of formulating a self-consistent definition of the entropy. In the present framework that obstacle does not arise: the entropy is a constitutive function whose state dependence is determined by the Liu procedure applied to the entropy inequality, see \re{eq:Liu-1}--\re{eq:Liu-8}, and the field equation emerges.
}

Holography implies that the balance of linear momentum \re{eq:bal-v} simplifies to
\begin{align}\label{holoevol_eq}
	\dot \qqv &= - \nabla \Phi,
\end{align}
therefore, the motion of the fluid can be described via a local, mass point like Newton equation. 

The second term of \re{eq:holop-1} is already in the form of the desired scalar potential, therefore, all we have to do is to reformulate the pressure gradient in the required form. This can be achieved easily through the Gibbs--Duhem relation
\begin{align}
	\label{eq:s-GD}
	\dd p \qf = \qrho s \dd T + \qrho \dd \mu \qf, 
\end{align}
which is a straightforward consequence of the Gibbs relation \re{eq:s-Gibbs} and the Euler relation \re{eq:s-Euler}. Finally, \re{eq:holop-1} can be given as
\begin{align}
	\label{eq:holopmu}
	\nabla \cdot \qqP \id = \qrho \nabla \left( \mu \qf + \frac{\delta \he}{\delta \qrho} \right) + \qrho s \nabla T.
\end{align}
Introducing the specific fluid enthalpy 
\begin{align}
    \label{eq:h-Euler}
	h \qf = u + \frac{p \qf}{\qrho} = T s + \mu \qf,
\end{align}
equation \re{eq:holopmu} can be transformed equivalently to
\begin{align}
	\label{eq:holoph}
	\nabla \cdot \qqP \id = \qrho \nabla \left( h \qf + \frac{\delta \he}{\delta \qrho} \right) - \qrho T \nabla s.
\end{align}

{Two remarks are in order before the scalar potential is identified. First, \re{eq:holopmu} and its enthalpy form \re{eq:holoph} below are not two different results but one and the same force written in two thermodynamic representations. Using the enthalpy representation of the Euler relation \re{eq:h-Euler} one has $\nabla h\qf = \nabla\left(Ts\right) + \nabla \mu\qf$, whence, via the Gibbs-Duhem relation \re{eq:s-GD},
\begin{align}
    \label{eq:tworep}
    \frac{1}{\qrho}\nabla p\qf = \nabla \mu\qf + s \nabla T = \nabla h\qf - T \nabla s.
\end{align}
The two representations differ by the exact gradient $\nabla\left(Ts\right)$, which is simply shuffled between the potential and the residual term. Consequently the residue itself is representation dependent---it is $-s\nabla T$ in the $\left(T,\mu\qf\right)$ representation and $+T\nabla s$ in the $\left(s,h\qf\right)$ representation---while its non-integrable content, \ie its curl, is not. Indeed, taking the curl of \re{eq:tworep} yields
\begin{align}
    \label{eq:integr}
    \nabla \times \left( \frac{1}{\qrho}\nabla p\qf \right) = \nabla s \times \nabla T,
\end{align}
in both cases. The necessary and sufficient condition for the specific pressure force $ \frac{1}{\qrho} \nabla \cdot \qqP \id $ to admit a scalar-potential representation is therefore
\begin{align}
    \label{eq:integrcond}
    \nabla s \times \nabla T = \zero,
\end{align}
that is, coinciding level surfaces of temperature and specific entropy, $T = T(s)$. Homothermal processes, $\nabla T = \zero$, and homoentropic ones, $\nabla s = \zero$, are the two familiar special cases; the general case $\nabla s \parallel \nabla T$ is a third. The condition is local: on a simply connected domain a closed one-form is exact, and the corresponding global question is of the same topological character as the single-valuedness discussed in Sec.~\ref{app:com-rep}.

Second, and more importantly for the purpose of this paper, one should ask what the weakly nonlocal extension contributes to the obstruction. The answer is: nothing. The whole of it---the Korteweg, quantum and gravitational contributions, all encoded in $ \frac{\delta\he}{\delta\qrho} $ in \re{eq:holop-1}---is of gradient form by construction and therefore obstructs the scalar-potential representation not at all. This is not obvious in advance. A pressure tensor containing $\nabla\otimes\nabla\qrho$ and $\nabla\otimes\nabla\qphi$ terms could perfectly well have produced a non-integrable residue, and that it does not is a genuine consequence of the Liu procedure. The only obstruction to classical holography with internal perfectness is the classical thermal one of \re{eq:integrcond}, which is already present for an ordinary Euler fluid. Weak nonlocality, in other words, does not compromise the holographic property; this is the central structural result of the present section.

Under the condition \re{eq:integrcond}, by introducing the scalar potential either in the $\left(T,\mu\qf\right)$ representation for homothermal processes as
\begin{align}
	\label{eq:sc-pot}
	\Phi &= \mu \qf + \pder{ \he }{\qrho} - \nabla \cdot \ppder{ \he}{\nabla \qrho},
\end{align}
or in the $\left(s,h\qf\right)$ representation for homoentropic processes as
\begin{align}
	\label{eq:sc-pot-h}
	\Phi &= h \qf + \pder{ \he }{\qrho} - \nabla \cdot \ppder{ \he}{\nabla \qrho},
\end{align} }%
the holographic momentum balance becomes identical to \re{holoevol_eq}. Consequently, the momentum balance is reduced to a mass point form, where the field effects are hidden in the density-dependent force field. This behavior is not exceptional in fluid mechanics, moreover, remarkable similarity with quantum mechanics have been demonstrated in several classical systems \cite{Bus15a,CasEta24b}.

 In general, the expressions \re{eq:sc-pot} and \re{eq:sc-pot-h} represent the non-thermal, constitutively determined part of the potential and remain well defined independently of the condition \re{eq:integrcond}. In the general integrable case $ \nabla s \times \nabla T = \zero $, the complete scalar potential is obtained by adding to \re{eq:sc-pot} a local function $ F_T^{} $ satisfying $ \dd F_T^{} = s \dd T $, or, equivalently, by subtracting from \re{eq:sc-pot-h} a local function $ F_s $ satisfying $ \dd F_s = T \dd s $. The forms given in \re{eq:sc-pot} and \re{eq:sc-pot-h} are recovered in the homothermal and homoentropic cases, respectively. If the integrability condition \re{eq:integrcond} fails, these expressions still represent the potential part of the specific force, but no scalar potential exists for the complete momentum balance. In the $\left(s,h\qf\right)$ representation, the remaining non-potential contribution is identified as the continuum entropic force, discussed in the next subsection.

{
\subsection{The residual force as a continuum entropic force}
\label{sec:entrf}

The structure uncovered above deserves a name, because it is one that already exists in physics. Substituting \re{eq:holoph} into the balance of linear momentum \re{eq:bal-v}, the specific force acting on a mechanically and internally perfect fluid is
\begin{align}
    \label{eq:entrforce}
    \dot{\qqv} = - \nabla \left( h\qf + \frac{\delta \he}{\delta \qrho} \right) + T \nabla s,
\end{align}
and the residual, non-potential term is $T\nabla s$. This is precisely the structure of an \emph{entropic force}, $\mathbf{F} = T\nabla S$: a force of purely thermodynamic origin, proportional to the temperature and to a gradient of entropy, with no underlying mechanical potential. Entropic forces are familiar from polymer elasticity, where the restoring force of an ideal chain is entirely entropic, from depletion interactions, and, in a much-discussed extrapolation, from proposals that gravity itself is of entropic origin~\cite{Ver11a}.

The observation we wish to record is that these familiar entropic forces are the \emph{integrable special case} of \re{eq:entrforce}. The classical examples are isothermal: for $\nabla T = \zero$ one has $T\nabla s = \nabla\left(Ts\right)$, the entropic force is a gradient, and it is absorbed into the potential without residue---which is exactly why it can be treated by equilibrium free-energy arguments. The general continuum entropic force is not a gradient, and the condition for it to be representable as a potential is the integrability condition \re{eq:integrcond} obtained above,
\begin{align}
    \nabla \times \left( T \nabla s \right) = \nabla T \times \nabla s = \zero.
\end{align}
Classical holography and the potential representability of the entropic force are thus one and the same requirement. When it fails, the fluid is not holographic, the momentum balance cannot be reduced to a mass-point Newton equation, and, as shown in Sec.~\ref{sec:vort}, vorticity is generated: $\nabla s \times \nabla T$ is also the classical baroclinic source term, so that the thermodynamic condition for holography coincides with the absence of baroclinic vorticity generation. We regard the last statement as a coherence check rather than as a new result, the baroclinic identity itself being classical, but the coincidence is worth noting because it links the thermodynamic and the kinematic obstructions.

Two lines of previous work should be mentioned. Treumann and Baumjohann~\cite{TreBau19a,BauTre23a} introduced a classical entropic force arising from the spatial inhomogeneity of disorder and applied it, in macroscopic variables, to magnetohydrodynamics; in their treatment $T\,\dd S(\qqr)$ is taken to be a mechanical potential. The present analysis is complementary: it gives the necessary and sufficient condition under which it is one, and obtains the force from the Second Law rather than postulating it. Separately, Rub\'i and coworkers~\cite{RegEta05a} have embedded entropic forces systematically in a nonequilibrium-thermodynamic setting at the mesoscopic level, through entropic potentials and entropic barriers; the continuum counterpart, with the corresponding thermodynamic validity conditions, appears not to have been given. A variational route to entropic forces in the continuum exists in the Lagrangian formulation of nonequilibrium thermodynamics~\cite{GayYos17a}, which provides an instructive contrast with the present one, where the Second Law replaces the variational principle throughout.

Finally, let us recall that for classical holography we required the convenient condition of internal perfectness. However, if the last term of equation \re{eq:holop} does not vanish, leads to an additional bulk force term of thermodynamic origin beyond the entropic force. }

{\subsection{Vorticity transport and preservation of irrotationality}}
\label{sec:vort}

Further properties of holographic fluids are revealed through the transformations of the momentum balance \re{holoevol_eq} into the following form
\begin{align}
	\label{eq:bal-v-vort}
	\pdt{\qqv} = - \nabla \left( \Phi + \frac{1}{2} \qqv \cdot \qqv \right) + \qqv \times \Tensor{\omega},
\end{align}
where $ \Tensor{\omega} = \nabla \times \qqv $ is the vorticity and the Lamb identity 
\begin{align}
	\qqv \cdot \nabla \qqv = \nabla \left( \frac{1}{2} \qqv \cdot \qqv \right) - \qqv \times \nabla \times \qqv 
\end{align}
was exploited.

There are two consequences. First, forming the curl of \re{eq:bal-v-vort} one obtains
\begin{align}
	\label{eq:bal-om}
	\dot{\Tensor{\omega}} - \Tensor{\omega} \cdot \nabla \qqv + \left( \nabla \cdot \qqv \right) \Tensor{\omega} = 0.
\end{align}
The last two terms on the l.h.s. of \re{eq:bal-om} characterize vortex stretching, the former is related to the deformation of the fluid element, while the latter one to its compressibility. {
Furthermore, \re{eq:bal-om} can be expressed in terms of the upper-convected time derivative---a.k.a. the Oldroyd deivative---, which for a vector field $ \ttensor{a} $ is defined as
\begin{align}
	\overset{\triangledown}{\ttensor{a}} = \dot{\ttensor{a}} - \ttensor{a} \cdot \nabla \qqv,
\end{align}
providing a frame-invariant material time rate of change within a coordinate system that translates, rotates, and deforms locally alongside the fluid element. Accordingly, \re{eq:bal-om} is the upper convected time derivative of the vorticity generalized by a compressibility related term \cite{Mor10a}, \ie
\begin{align}\label{vortLie_eq_0}
	\overset{\triangledown}{\Tensor{\omega}} + \left( \nabla \cdot \qqv \right) \Tensor{\omega} = 0. 
\end{align}
Utilizing the mass continuity equation \re{eq:bal-m} equation \re{vortLie_eq_0} yields
\begin{align}\label{vortLie_eq}
	\qrho \left(\frac{\Tensor{\omega}}{\qrho}\right) \!\!\!\overset{\triangledown}{\vphantom{\frac{\Tensor{\omega}}{\qrho}}}=0,
\end{align}
}
{
expressing the geometric concept of vortex freeze-in through the specific vorticity $ \frac{\Tensor{\omega}}{\qrho} $. This specific vorticity field, measured in $ \rm \frac{m^3}{kg \, s} = \frac{m^2/s}{kg/m} $, can be physically interpreted as the local circulation per unit mass density along a line element. Consequently, any fluid stretching or compression dynamically scales the vorticity magnitude to strictly preserve the total circulation along any closed material loop.}

{
Provided that a sufficiently smooth solution of the governing fluid equations exists, \re{vortLie_eq} preserves the irrotational condition: if the fluid is initially rotation free, \ie $ \Tensor{\omega} ( 0, \qqr ) = \zero $, then $ \Tensor{\omega} ( t, \qqr ) = \zero $ throughout the subsequent evolution.
}
In this case, there exists a velocity potential $ \qS$, defined as $\qqv = \nabla \qS $ and \re{eq:bal-v-vort} simplifies to a Bernoulli equation:
\begin{align}
	\label{eq:Bern}
	\pdt{\qS} + \frac{1}{2} \nabla \qS \cdot \nabla \qS + \Phi = \rm{const}.
\end{align}
Considering integrable, homothermal or homoentropic processes, evolution equations of perfect fluids are reduced to the mass continuity equation \re{eq:bal-m} and the Bernoulli equation \re{eq:Bern}.

\subsection{Wave function representation of perfect fluids} \label{app:com-rep}

In the previous sections, we presented how classical holographic property simplifies the evolution equations of ideal fluids, and for integrable, homothermal or homoentropic processes of perfect fluids two scalar governing equations emerged: are the mass continuity equation and the Bernoulli equation. In this case one can combine the density and velocity potential fields into a single complex field, a wave function, expressed in polar form as
\begin{align}
	\label{eq:wave-fun}
	\Psi \left( t, \qqr \right) = R \left( t, \qqr \right) \ee^{ \ii \frac{\qS \left( t, \qqr \right)}{\qS_0} }
\end{align}
with the imaginary unit $ \ii = \sqrt{-1} $, $ R = \sqrt{\qrho} $ {for strictly positive density $ \qrho > 0 $,} and the arbitrary non-zero constant $ \qS_0$. Expressing the mass continuity equation \re{eq:bal-m} via $ R $ and $ \qS $, then considering the linear combination of the mass continuity equation and the Bernoulli equation with the---so far---unknown functions $\mathsf{f}_1 $ and $\mathsf{f}_2 $ one obtains
\begin{align}
	\label{eq:Sch-Mad-inv}
	2 R \mathsf{f}_1 \left( \pdt{R} + \nabla R \cdot \nabla \qS + \frac{1}{2} R \Delta \qS \right) + \mathsf{f}_2 \left( \pdt{\qS} + \frac{1}{2} \nabla \qS \cdot \nabla \qS + \Phi \right) = 0.
\end{align}
This equation can be reformulated in terms of a single complex variable, namely the wave function \re{eq:wave-fun}, provided that $ \mathsf{f}_1 = \frac{1}{2 R} \ee^{\ii \frac{\qS}{\qS_0}} $ and $ \mathsf{f}_2 = \ii \frac{R}{\qS_0} \ee^{\ii \frac{\qS}{\qS_0}} $, therefore,
\begin{align}\label{eq:protoSch}
	\ii \qS_0 \pdt{\Psi} = \left( - \frac{\qS_0^2}{2} \Delta + \left( \Phi + \frac{\qS_0^2}{2} \frac{\Delta | \Psi |}{| \Psi |} \right) \right) \Psi 
\end{align}
is equivalent to \re{eq:Sch-Mad-inv}.

This unified complex evolution equation becomes linear, if the holographic potential is chosen properly. Accordingly, it is straightforward to define the holographic potential as
\begin{align}\label{def:protoSch-free}
	\Phi_{\rm free} = - \frac{\qS_0^2}{2} \frac{\Delta | \Psi |}{| \Psi |},
\end{align}
which leads to the linear partial differential equation
\begin{align}
	\label{eq:preSchrod-free}
	\ii \qS_0 \pdt{\Psi} = - \frac{\qS_0^2}{2} \Delta \Psi.
\end{align}
Finally, the choice $\qS_0 = \frac{\hbar}{m}$ results in the free-particle Schrödinger equation
\begin{align}
	\label{eq:Schrod-free}
	\ii \hbar \pdt \Psi = - \frac{\hbar^2}{2 m} \Delta \Psi 
\end{align}
with particle mass $ m $ and reduced Planck constant $ \hbar $. Therefore, the Bohmian term $- \frac{\hbar^2}{2 m^2} \frac{\Delta | \Psi |}{| \Psi |} $ represents the self-interaction of the fluid-particle. Bounded states can be formulated via external effects, which can be also included adding a further term to the holographic potential, \ie
\begin{align}\label{def:protoSch}
	\Phi_{\rm bounded} = - \frac{\hbar^2}{2 m^2} \frac{\Delta | \Psi |}{| \Psi |} + \frac{V}{m} = \frac{1}{m} \left( - \frac{\hbar^2}{2 m} \frac{\Delta | \Psi |}{| \Psi |} + V \right),
\end{align}
where the potential energy function $ V $ does not depend on the fluid properties, it characterizes the external field. The transition from a continuum to a particle perspective becomes evident when the potential energy $V$ is used instead of its mass specific equivalent $\frac{V}{m}$. Summarizing, a Schrödinger-like equation via a general holographic potential (without external fields) can be formulated as
\begin{align}
	\label{eq:Sch-Bohm}
	\ii \hbar \pdt{\Psi} = \left( - \frac{\hbar^2}{2 m} \Delta + \left( m \Phi + \frac{\hbar^2}{2 m} \frac{\Delta | \Psi |}{| \Psi |} + V \right) \right) \Psi.
\end{align}

In the inverse way presented hydrodynamic formulation of quantum mechanics---usually called the Bohmian reformulation---is considered as a helpful tool to prove existence and stability theorems \cite{CarEta12a}. 

{
The admissible domain of the representation is the open set on which $\qrho > 0$. This excludes the nodes of the wave function and the cores of quantized vortices, where the quantum potential and the terms proportional to $\frac{1}{\qrho}$ appearing in Sec.~\ref{sec:spec}\,E---among them $ \frac{C}{\qrho} \Delta \phi$ and $\Delta \ln |\Psi|^2$---are singular. The exclusion is not merely technical: these are precisely the loci at which quantized circulation resides, and hence where the topological information encoded in the single-valuedness assumption is concentrated. Extending the treatment to them, in particular examining whether the dissipative terms of the present framework regularize the vortex core, is a mathematical and physical challenge.

It is important to distinguish the result obtained here from a derivation of quantum mechanics from thermodynamics. The Second Law constrains the admissible continuum structure, but it does not by itself determine the constant $\qS_0 = \frac{\hbar}{m}$, or provide the complete probabilistic and operator structure of quantum mechanics. From a thermodynamic point of view, the Schrödinger-like complex evolution equation above is already a constitutive specialisation, and one must choose an appropriate holographic potential to obtain the linear Schrödinger equation. This is the choice that yields the Bohm potential and the associated Fisher-form extra energy. Whether the thermodynamic background points to a deeper physical connection between fluid theories and quantum dynamics remains an open question.  
}

{\subsection{Logical hierarchy of the perfect-fluid results}}

\begin{figure*}[!t]
	\centering
	\includegraphics[width=\textwidth]{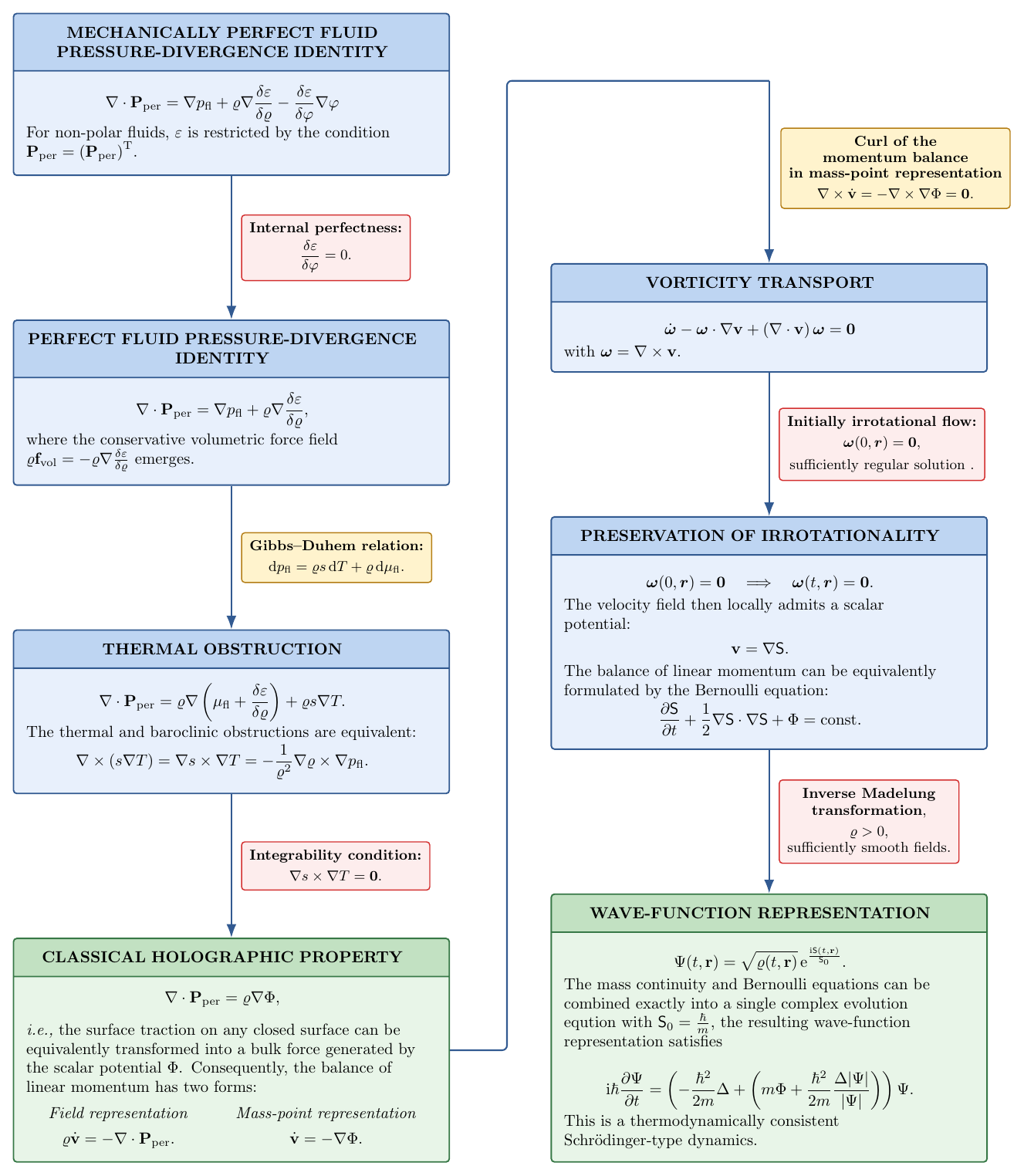}
	\caption{{The main steps leading to the wave function representation of perfect fluids. Conditions and thermodynamic consequences are denoted by the red and yellow boxes, respectively, while the main results are in the green boxes.}}
	\label{fig:perfect-fluid-hierarchy}
\end{figure*}

{
The logical hierarchy of the results derived in this section is summarized in Fig.~\ref{fig:perfect-fluid-hierarchy}. Mechanical perfectness yields the general pressure-divergence identity, while internal perfectness and the integrability of the thermal contribution lead successively to the classical holographic property. The vorticity transport equation, the preservation of initially irrotational flow, and the wave-function representation then follow as further conditional consequences under the additional assumptions indicated in the figure.
}

\section{Special cases} \label{sec:spec}

Below we discuss some well-known physical systems. In all cases, we assume, that the \adden \ density is quadratic in the gradients and given as
\begin{align}
	\label{eq:quad-en}
	\he \left( \qrho, \nabla \qrho, \qphi, \nabla \qphi \right) =
	a_1 ( \qrho ) \qphi + \frac{a_2 ( \qrho )}{2} \qphi^2 + \frac{1}{2}
	\begin{pmatrix}
		\nabla \qrho & \nabla \qphi
	\end{pmatrix} \cdot
	\begin{pmatrix}
		b_{1} ( \qrho ) & b_{12} ( \qrho ) \\
		b_{12} ( \qrho ) & b_{2} ( \qrho )
	\end{pmatrix}
	\begin{pmatrix}
		\nabla \qrho \\ \nabla \qphi
	\end{pmatrix} 
\end{align}

with the density-dependent functions $ a_1 $, $ a_2 $, $ b_1 $, $ b_{12} $ and $ b_2 $. 

The positive semidefiniteness of the coefficient matrix  
\begin{align}
    \label{eq:Bmatrix}
    \ttensor{B}(\qrho) = \begin{pmatrix} b_1(\qrho) & b_{12}(\qrho) \\ b_{12}(\qrho) & b_2(\qrho) \end{pmatrix}
\end{align}
is a particular form of the usual conditions of thermodynamic stability, related to the concavity of the entropy, but in the gradient variables. All
models collected in Table~\ref{tab} are thermodynamically compatible; some of them
satisfy the convexity of the gradient block only in a restricted range of densities.
The precise dynamical role of \re{eq:Bmatrix} --- it governs the short-wavelength
behaviour of the linearized equations --- is identified in Sec.~\ref{sec:spec}\,D.
Convexity jointly in all variables is not demanded either: already the interaction
term $a_1(\qrho)\qphi$ of the gravitational case is not jointly convex in
$\left(\qrho,\qphi\right)$, and it is precisely this term that carries the classical
Jeans instability, as discussed below.
Note that the individual density dependence of $ \he $ is excluded since this can be included in the fluid internal energy $ u \qf $.

The general---not necessarily symmetric---pressure tensor \re{eq:P_id} is calculated from \re{eq:quad-en}, \ie in index notation
\begin{align}
	\nonumber
	\left( \qP \id \right)_i^j &= \left( p \qf -\he + \qrho \pder{\he}{\qrho} \right) \delta_i^j - \frac{\qrho^2}{2} \pd_k \left( \frac{1}{\qrho}\ppder{\he}{\pd_l \qrho} \right) \left( \delta_l^k \delta_i^j + \delta_l^j \delta_i^k \right) + 
	\ppder{\he}{\pd_j \qphi} \pd_i \qphi \\
	\nonumber
	& = \bigg( p \qf + (\qrho a_1' -a_1)\qphi + \frac{1}{2} \left( \qrho a_2' - a_2 \right) \qphi^2 + \frac{1}{2} \left( \qrho b_{12}' - b_{12} \right) \pd^k \qrho \pd_k \qphi +
	\frac{1}{2} (\qrho b_2' -b_2) \pd^k \qphi \pd_k \qphi \\
	\nonumber
	& \hskip 2.7ex- 
	\frac{\qrho b_{1} }{2} \pd_k^k \qrho - 
	\frac{\qrho b_{12} }{2} \pd_k^k \qphi \bigg) \delta_i^j - \frac{1}{2} \left( \left( \qrho b_{1}' - b_{1} \right) \pd_i \qrho \pd^j \qrho + \qrho b_{1} \pd_i^j \qrho + \qrho b_{12} \pd_i^j \qphi - 2 b_{2} \pd_i \qphi \pd^j \qphi \right) \\
	\label{eq:quad-en-p}
	& \hskip 2.7ex - \frac{1}{2} \left( \qrho b_{12}' - b_{12} \right) \pd_i \qrho \pd^j \qphi + b_{12} \pd_i \qphi \pd^j \qrho
\end{align}
where $ ' $ denotes the derivative w.r.t.~$ \qrho $. The pressure tensor in the invariant notation is written as
\begin{align}
	\nonumber
	\qqP \id & = \bigg( p \qf + (\qrho a_1'-a_1) \qphi + 
	\frac{1}{2} \left( \qrho a_2' - a_2 \right) \qphi^2 + 
	\frac{1}{2}\left( \qrho b_{12}' - b_{12} \right) \nabla \qrho \cdot \nabla \qphi + 
	\frac{1}{2}\left( \qrho b_{2}' - b_{2} \right) \nabla \qphi \cdot \nabla \qphi \\
	\nonumber
	& \hskip 2.7ex
	- \frac{\qrho b_{1} }{2} \Delta \qrho - \frac{\qrho b_{12}}{2} \Delta \qphi \bigg) \ident - \frac{1}{2} \big( \left( \qrho b_{1}' - 
	b_{1} \right) \nabla \qrho \otimes \nabla \qrho + \qrho b_{1} \nabla \otimes \nabla \qrho + 
	\qrho b_{12}\nabla \otimes \nabla \qphi - 2 b_{2} \nabla \qphi \otimes \nabla \qphi \big) \\
	\label{eq:quad_id_p}
	& \hskip 2.7ex
	- \frac{1}{2}\left( \qrho b_{12}' - b_{12} \right)\nabla \qrho \otimes \nabla \qphi + b_{12}
	\nabla \qphi \otimes \nabla \qrho,
\end{align}
where $ \Delta = \nabla \cdot \nabla $ is the Laplace operator. Note that only the last two terms are not symmetric in \re{eq:quad-en-p}. Separating the symmetric and skew-symmetric parts in these terms one obtains
\begin{align}
	\nonumber
	b_{12} \pd_i \qphi \pd^j \qrho &- \frac{1}{2} \left( \qrho b_{12}' - b_{12} \right) \pd_i \qrho \pd^j \qphi \\
	& \hskip 3ex = \frac{1}{4} \left( 3 b_{12} - \qrho b_{12}' \right) \left( \pd_i \qphi \pd^j \qrho + \pd_i \qrho \pd^j \qphi \right) + \frac{1}{4} \left( \qrho b_{12}' + b_{12} \right) \left( \pd_i \qphi \pd^j \qrho - \pd_i \qrho \pd^j \qphi \right),
\end{align}
therefore, the only opportunity to obtain a symmetric pressure tensor is
\begin{align}\label{symcond_eq}
	\qrho b_{12}' + b_{12} = 0,
\end{align}
from which follows that $ b_{12} (\qrho) = \frac{C}{\qrho} $, where $ C $ is a constant. Summarizing, the symmetric pressure tensor compatible with the quadratic \adden \ density contribution is
\begin{align}
	\nonumber
	\qqP \id & = \bigg( p \qf + (\qrho a_1'-a_1) \qphi + 
	\frac{1}{2} \left( \qrho a_2' - a_2 \right) \qphi^2 - \frac{C}{\qrho} \nabla \qrho \cdot \nabla \qphi + 
	\frac{1}{2}\left( \qrho b_{2}' - b_{2} \right) \nabla \qphi \cdot \nabla \qphi - \frac{\qrho b_{1} }{2} \Delta \qrho - \frac{C}{2} \Delta \qphi \bigg) \ident \\
	\nonumber
	& \hskip 2.7ex
	- \frac{1}{2} \big( \left( \qrho b_{1}' - 
	b_{1} \right) \nabla \qrho \otimes \nabla \qrho + \qrho b_{1} \nabla \otimes \nabla \qrho + 
	C \nabla \otimes \nabla \qphi - 2 b_{2} \nabla \qphi \otimes \nabla \qphi \big) \\
	& \hskip 2.7ex
	+ \frac{C}{\qrho} \left( \nabla \qrho \otimes \nabla \qphi + \nabla \qphi \otimes \nabla \qrho \right).
\end{align}
Based on \re{eq:sc-pot}, the corresponding scalar potential is
\begin{align}
	\label{eq:quad_pot}
	\Phi = \mu \qf + a_1' \qphi + \frac{a_2'}{2} \qphi^2 - \frac{1}{2} b_1' \nabla \qrho \cdot \nabla \qrho
	+ \frac{b_{2}'}{2} \nabla \qphi \cdot \nabla \qphi  
	- b_1 \Delta \qrho - \frac{C}{\qrho} \Delta \qphi.
\end{align}
The field equation of the scalar internal variable field \re{eq:GL} becomes
\begin{align}
	\label{eq:quad_field}
	a_1 + a_2 \qphi - b_2 \Delta\qphi - b_2' \nabla \qrho \cdot \nabla \qphi - b_{12}' \nabla \qrho \cdot \nabla \qrho - b_{12} \Delta \qrho =0.
\end{align}

Table \ref{tab}.~presents some important perfect fluid models, their relationship to \adden \ density \re{eq:quad-en}, as well as their pressure tensors and the corresponding potentials. All these cases are determined solely by the appropriate form of their \adden\ functions, $\he(\qrho,\nabla\qrho,\qphi,\nabla\qphi)$. {In the analysis of each special case, we provide the corresponding forms of both the balance of linear momentum---\ie the equation of motion---and the scalar-field equation obtained from \re{eq:quad_field}. The latter is the specific realization of \re{eq:GL} associated with internal perfectness.}

\begin{table}[h!]
	\begin{minipage}[t]{0.15\textwidth}
		\begin{raggedright}
			\textbf{Euler fluids}
			\vskip 15ex
			\textbf{Korteweg-like fluids} (including superfluid models and Schrödinger--Madelung fluids)
			\vskip 12.45ex
			\textbf{Newtonian gravity}
			\vskip 20ex
			\textbf{Modified Newtonian gravity}
			\vskip 27ex
			\textbf{Generalized Schrödinger--Newton equations}
			\vskip 1ex ~
		\end{raggedright}
	\end{minipage}
	\hfill
	\begin{minipage}[t]{0.83\textwidth}
		\vskip -4.8ex
		\begin{align}
			a_1 &= 0, \ a_2 = 0, \ b_{1} = 0, \ b_{12} = 0, \ b_2 = 0, \\
			\he_{\rm E}^{} &\equiv 0, \label{ex_Euler}\\
			\qqP_{\rm E}^{} &= p \qf, \\
			\Phi_{\rm E}^{} &= \mu \qf. \\ \nonumber \\
			a_1 &= 0, \ a_2 = 0, \ b_{1} \neq 0, \ b_{12} = 0, \ b_2 = 0, \\
			\he_{\rm K}^{} &= \frac{b_1}{2} \nabla \qrho \cdot \nabla \qrho, \label{ex_Korteweg}\\
			\qqP_{\rm K}^{} &= \left( p \qf - \frac{\qrho b_1}{2} \Delta \qrho \right)\ident - \frac{1}{2} \left( \qrho b_1' - b_1 \right) \nabla \qrho \otimes \nabla \qrho - \frac{\qrho b_1}{2} \nabla \otimes \nabla \qrho, \\
			\Phi_{\rm K}^{} &= \mu \qf - \frac{1}{2} b_1' \nabla \qrho \cdot \nabla \qrho - b_1 \Delta \qrho.\\ \nonumber \\
			\qphi &\equiv \phi, \ a_1 = \qrho, \ a_2 = 0, \ b_{1} = 0, \ b_{12} = 0, \ b_2 = \frac{1}{4 \pi G}, \ie \\
			\he_{\rm N}^{} &= \qrho \phi + \frac{1}{8 \pi G} \nabla \phi \cdot \nabla \phi, \label{ex_Newton}\\
			\qqP_{\rm N}^{} &= \left( p \qf - \frac{1}{8 \pi G} \nabla \phi \cdot \nabla \phi \right) \ident + \frac{1}{4 \pi G} \nabla \phi \otimes \nabla \phi, \\
			\Phi_{\rm N}^{} &= \mu \qf + \phi. \\ \nonumber \\
			\qphi &\equiv \phi, \ a_1 = \qrho, \ a_2 = 0, \ b_{1} = 0, \ b_{12} = \frac{C}{\qrho}, \ b_2 = \frac{1}{4 \pi G}, \ie \\
			\he_{\rm mN}^{} &= \qrho \phi + \frac{C}{\qrho} \nabla \qrho \cdot \nabla \phi + \frac{1}{8 \pi G} \nabla \phi \cdot \nabla \phi,\label{ex_modNewton} \\
			\nonumber
			\qqP_{\rm mN}^{} &= \left( p \qf - \frac{C}{\qrho} \nabla \qrho \cdot \nabla \phi - \frac{1}{8 \pi G} \nabla \phi \cdot \nabla \phi - \frac{C}{2} \Delta \phi \right) \ident - \frac{C}{2} \nabla \otimes \nabla \phi, \\
			&\hskip 3ex + \frac{1}{4 \pi G} \nabla \phi \otimes \nabla \phi + \frac{C}{\qrho} \left( \nabla \qrho \otimes \nabla \phi + \nabla \phi \otimes \nabla \qrho \right), \\
			\Phi_{\rm mN}^{} &= \mu \qf + \phi - \frac{C}{\qrho} \Delta \phi. \\ \nonumber \\
			\qphi &\equiv \phi, \ a_1 = \qrho, \ a_2 = 0, \ b_{1} = \frac{\hbar^2}{4 m^2 \qrho}, \ b_{12} = \frac{C}{\qrho}, \ b_2 = \frac{1}{4 \pi G}, \\
			\he_{\rm SN}^{} &= \qrho \phi + \frac{\hbar^2}{8 m^2 \qrho} \nabla \qrho \cdot \nabla \qrho + \frac{C}{\qrho} \nabla \qrho \cdot \nabla \phi + \frac{1}{8 \pi G} \nabla \phi \cdot \nabla \phi, \label{ex_SchNewton}\\
			\nonumber
			\qqP_{\rm SN}^{} &= \bigg( p \qf - \frac{C}{\qrho} \nabla \qrho \cdot \nabla \phi -	\frac{1}{8 \pi G} \nabla \phi \cdot \nabla \phi - \frac{\hbar^2}{8 m^2} \Delta \qrho - \frac{C}{2} \Delta \phi \bigg) \ident + \frac{\hbar^2}{4 m^2 \qrho} \nabla \qrho \otimes \nabla \qrho, \\
			& \hskip 3ex - \frac{\hbar^2}{8 m^2} \nabla \otimes \nabla \qrho - \frac{C}{2} \nabla \otimes \nabla \phi + \frac{1}{4 \pi G} \nabla \phi \otimes \nabla \phi + \frac{C}{\qrho} \left( \nabla \qrho \otimes \nabla \phi + \nabla \phi \otimes \nabla \qrho \right), \\
			\Phi_{\rm SN}^{} &= \mu \qf + \phi - \frac{\hbar^2}{4 m^2} \left( \frac{\Delta \qrho}{\qrho} - \frac{\nabla \qrho \cdot \nabla \qrho}{2 \qrho^2} \right) - \frac{C}{\qrho} \Delta \phi = \mu \qf + \phi - \frac{\hbar^2}{2 m^2} \frac{\Delta \sqrt{\qrho}}{\sqrt{\qrho}} - \frac{C}{\qrho} \Delta \phi.
		\end{align}
	\end{minipage}
	\caption{Distinguished homothermal perfect fluid models compatible with the quasi-quadratic \adden \ density \re{eq:quad-en}.}
	\label{tab}
\end{table}

\subsection{Euler fluids}

Assuming a zero \adden \ density contribution \re{ex_Euler}, Euler fluids are obtained, then holographic property leads to the reformulation of the Friedmann form of the Euler flow equations, which for homothermal or homoentropic processes can be given as
\begin{align}
	\dot \qqv = - \nabla \mu\qf 
	\quad\text{or} \quad 
	\dot \qqv = - \nabla h\qf.
\end{align}
This is the homothermal or homoentropic special case of a relation that Friedmann obtained in 1922 for a perfect gas~\cite{Fri22b} and that is now generally known as Crocco's theorem~\cite{Cro37a}. In the general case the same manipulation yields  

\begin{align}
    \pdt{\qqv} + \nabla \left( h\qf + \frac{1}{2} \qqv \cdot \qqv \right) + \Tensor{\omega} \times \qqv - T \nabla s = \zero
\end{align}
whose non-potential term $T\nabla s$ is precisely the residual entropic force discussed in Sec.~\ref{sec:entrf}. Friedmann's priority is not widely known, the result being usually attributed to Crocco alone.

Accordingly, for any Euler fluid, the chemical potential (or equivalently the specific enthalpy) is a particular mechanical potential at the same time. In contrast to the usual interpretation of potentials, this potential expresses material properties, not the effect of an external field, therefore, to solve this modified Friedmann equation, it must be coupled to the balance equations of mass and energy, because $ \mu\qf $ and $h\qf$ are density and temperature dependent. 

\subsection{Korteweg-like fluids} \label{subsec:kor}

Korteweg fluids are characterized by the pressure tensor
\begin{align}
	\label{eq:Kor-fl} 
	\qqP_{\rm Kor} & = \left( p \qf - \alpha_1 \Delta \qrho - \alpha_2 ( \nabla \qrho )^2 \right) \ident - \alpha_3 \nabla \qrho \otimes \nabla \qrho - \alpha_4 \nabla \otimes \nabla \qrho 
\end{align}
with the density and temperature dependent coefficients $ \alpha_1 $, $ \alpha_2 $, $ \alpha_3 $ and $ \alpha_4 $. If all coefficient functions in \re{eq:quad-en} are zero except for $ b_{1} $, then a Korteweg-like fluid is obtained with the \adden \ function \re{ex_Korteweg}. To be more precise, the special Korteweg fluid with the coefficients $ \alpha_1 = \alpha_4 = \frac{\qrho b_1}{2} $, $ \alpha_2 = 0 $ and $ \alpha_3 = \frac{1}{2}(b_1-\qrho b_1')$ is resulted. Although this model is a strongly restricted Korteweg fluid, it paves the way for the hydrodynamic formulation of quantum phenomena.

Now, if $ b_1 $ is identified with $ \frac{1}{\qrho} \frac{\hbar^2}{4m^2} $, then the corresponding mass-specific \adden\ (the \adden\ density divided by the mass density),
\begin{align}
	\label{eq:Fisher} 
	\frac{\varepsilon_{\rm K}}{\qrho} &=
	\frac{\hbar^2}{8 m^2} \frac{\nabla \qrho \cdot \nabla \qrho}{ \qrho^2},
\end{align}
connects to the Fisher information measure \cite{Fis59b}. But, at the same time, it has a remarkable physical property, too. Namely, it is additive and unique with the additivity property among first-order weakly nonlocal functions, like the logarithm in the case of local functions. Therefore, this energy form gives the only possibility that multicomponent fluids can represent independent particles in a probabilistic interpretation \cite{Van06a,Van23a}.

In this case, the obtained holographic potential can be given as the sum of the Bohm potential and a fluid chemical potential as
\begin{align}
	\Phi_{\rm K} &= \mu \qf - \frac{\hbar^2}{4m^2} \left( \frac{\Delta \qrho}{\qrho} - \frac{\nabla \qrho \cdot \nabla \qrho}{2 \qrho^2} \right) = \mu \qf - \frac{\hbar^2}{2 m^2} \frac{\Delta \sqrt{\qrho}}{\sqrt{\qrho}}.
\end{align}
If the temperature is homogeneous, the difference of $\Phi_{\rm K}$ and the general holographic potential $\Phi$ in \re{def:protoSch-free} is only the density dependent chemical potential $\mu\qf$, then the complex wave equation \re{eq:Sch-Bohm} reduces to
\begin{align}
	\ii \hbar \pdt{\Psi} = \left( - \frac{\hbar^2}{2 m} \Delta + \left( V + m \mu\qf \left( | \Psi |^2 \right) \right) \right) \Psi.
\end{align}
For particular choices of the chemical potential and the corresponding thermodynamic potential in polynomial and logarithmic forms, Gross--Pitaevskii, Ginzburg--Landau, Ginzburg--Sobyanin and Bia\l{}ynicky--Birula--Mycielski-like equations emerge for superfluids. Let us here remark that in the thermodynamic formulation the coefficient $ b_1 $ is a free {constitutive} parameter{---subject to the stability requirements appropriate to the selected physical interpretation,---}which opens further ways in interpretation of experimental data.

Assuming zero chemical potential and correspondingly zero fluid pressure, the Schrödinger equation is obtained, \ie the Schrödinger--Madelung fluid appears, characterized by the pressure tensor
\begin{align}
	\qqP_{\rm SM} & = \frac{\hbar^2}{4 m^2} \frac{\nabla \qrho \otimes \nabla \qrho}{\qrho} - \frac{\hbar^2}{8 m^2} \left( \Delta \qrho \ident + \nabla \otimes \nabla \qrho \right) \\
	\label{eq:P-SM} 
	&= - \frac{\hbar^2}{4 m^2} \qrho \nabla \otimes \frac{\nabla \qrho}{\qrho} - \frac{\hbar^2}{8 m^2} \left( \Delta \qrho \ident - \nabla \otimes \nabla \qrho \right).
\end{align}
The first term of \re{eq:P-SM} is exactly the quantum pressure derived from the Bohm potential,
\begin{align}
	\nonumber
	\qrho \pd_i \left( - \frac{\hbar^2}{4 m^2} \left( \frac{\pd_j^j \qrho}{\qrho} - \frac{\pd^j \qrho \pd_j \qrho}{2 \qrho^2} \right) \right) &= \pd_i \left( -\frac{\hbar^2}{4 m^2} \pd_j^j \qrho \right) + \frac{\hbar^2}{4 m^2} \frac{\pd_j^j \qrho \pd_i \qrho}{\qrho} + \frac{\hbar^2}{4 m^2} \frac{\pd^j_i \qrho \pd_j \qrho}{\qrho} - \frac{\hbar^2}{4 m^2} \frac{\pd^j \qrho \pd_j \qrho \pd_i \qrho}{\qrho^2} \\
	&= \pd_j \left( - \frac{\hbar^2}{4 m^2} \left( \pd^j_i \qrho - \frac{\pd_i \qrho \pd^j \qrho}{\qrho} \right) \right) = \pd_j \left( - \frac{\hbar^2}{4 m^2} \qrho \pd_i \frac{\pd^j \qrho}{\qrho} \right).
\end{align}
It is easy to prove that the second term of \re{eq:P-SM} is divergence free, therefore, the quantum pressure tensor is defined only up to a full-divergence term. This highlights that the thermodynamic methodology determines a distinguished pressure tensor, different from the classical one. This difference may be important in dissipative processes and can influence whether and under what conditions the processes of ideal fluid equations are attractors (or not) of the corresponding dissipative dynamics \cite{BreEta16a}.


\subsection{Newtonian gravity} \label{subsec:ng}

Identifying the internal variable via the gravitational potential $ \phi $, \ie $ \qphi \equiv \phi $, and assuming that all coefficients in \re{eq:quad-en} are zeros except for $ a_1 = \qrho $ and $ b_2 = \frac{1}{4 \pi G} $ (with the gravitational constant $ G $) one obtains the governing equations of Newtonian self-gravitating fluids. According to equation \re{eq:euler}, the conservative volumetric force field $ \ttensor{f}_{\rm vol} = - \nabla \frac{\delta \he}{\delta \qrho} = - \nabla \phi $ emerges in the balance of linear momentum, as a direct mechanical manifestation of the gravitational field. It is also remarkable, that the thermodynamically proposed gravitational \adden\ density is not the energy of the gravitational field alone, but the sum of the field and the interaction energies \cite{VanAbe22a}. {Expression \re{ex_Newton} is the Ohanian energy density of Newtonian gravity~\cite{OhaRuf13b}: the sum of the matter--field interaction term $\qrho\phi$ and the field term $\frac{1}{8\pi G}\nabla\phi\cdot\nabla\phi$, as distinguished from the Maxwell convention, which retains the field term alone, and the Bondi convention, which retains half the interaction term. All three yield the same integrated energy under suitable boundary conditions~\cite{Pet81a}, and it is the thermodynamic derivation that selects this particular combination; a detailed comparison of the three conventions and of the reason for this selection is given in~\cite{TraEta26a}.}

Thanks to the holographic property the coupling between the equation of motion and the field equation is directly obtained, \ie
\begin{align}
	\label{eq:NG-eq-mot}
	\dot \qqv &= - \nabla \left( \mu \qf + \phi \right), \\
	\label{eq:Poisson}
	\Delta \phi &= 4 \pi G \qrho.
\end{align}
Let us observe that in \re{eq:NG-eq-mot}, the gravitational potential is a mechanical potential as well, due to the holographic property, as it must be.

\subsection{Modified Newtonian gravity -- field equations with weakly nonlocal source terms}

With the previous assumptions on Newtonian gravity together with the density-gravity coupling term characterized by the unique coefficient $ b_{12} = \frac{C}{\qrho} $ one obtains the \adden \ density \re{ex_modNewton} as a starting point. Then the derived coupled motion and field equations
\begin{align}
	\label{eq:mNG-eq-mot}
	\dot \qqv &= - \nabla \left( \mu \qf + \phi - \frac{C}{\qrho} \Delta \phi \right), \\
	\label{eq:mPoisson}
	\Delta \phi &= 4 \pi G \left( \qrho - C \nabla \cdot \left( \frac{\nabla \qrho}{\qrho} \right) \right).
\end{align}

The resulting field equation \re{eq:mPoisson} is similar to the modified Poisson equation
\begin{align}
	\label{eq:poi-pal}
	\Delta \phi = \frac{\kappa}{2} \left( \qrho + 2 \alpha \Delta \qrho \right) 
\end{align}
(with the constants $ \alpha $ and $ \kappa $) obtained as the Newtonian limit of the Eddington-inspired Born-Infeld gravity \cite{BanFer10a} or Palatini formulation of $f(R)$ gravity \cite{TonEta20a}. The modified theory can be tested with geophysical data, \cite{KozEta25a,KozWoj26a}, and presumably with neutron star observations.

Our non-relativistic analysis highlights that according to the thermodynamic requirements, if the holographic potential changes, see the r.h.s. of \re{eq:mNG-eq-mot}, then it implies the change of the pressure tensor. {Note} that applying a constant coupling parameter $b_{12}$ our derived field equation has the same form as \re{eq:poi-pal}, but then a nonsymmetric pressure tensor appears. Investigation of this case requires repeating the whole analysis, also including the balance equation of angular momentum.

{
In the relativistic theories cited above, the density-gradient correction arises in the non-relativistic limit of the modified gravitational field equations. In the present approach, a related correction emerges instead from the density--gravity gradient contribution to the weakly nonlocal extra energy. Requiring a symmetric pressure tensor restricts the coupling to $ b_{12} = \frac{C}{\qrho} $, thereby replacing the density Laplacian in \re{eq:poi-pal} by the nonlinear expression $ \nabla \cdot \left( \frac{\nabla \qrho}{\qrho} \right) = \Delta \ln \qrho $. The resulting equation may therefore be regarded as a thermodynamically motivated counterpart of these modified Poisson equations, although the physical viability of the coupling and its compatibility with gravitational tests remain to be investigated.
}

{
A further mathematical issue arises in the present $ C \neq 0 $ specialization: the quadratic gradient block in \re{eq:quad-en} is indefinite with respect to independent variations of $ \nabla \qrho $ and $ \nabla \phi $. Accordingly, thermodynamic admissibility does not directly ensure stability, which requires an analysis of the complete constrained matter–field system.
}

Although we have neglected cross-effects {in the present analysis}, an antisymmetric coupling in \re{eq:Ons-1} and \re{eq:Ons-2}, \ie $ L_{\pi f} = - L_{f \pi} ${, leads to} a further, {consistent} modification of the Poisson equation of Newtonian gravity. {We do not investigate this Casimir-coupled case here; however, it was analyzed in a previous work, where the resulting modification was applied to galactic rotation curves and shown to provide a promising thermodynamic approach to dark-matter-like gravitational phenomena \cite{PszoVan24a}.}

{The admissibility of the coupling requires closer examination, since the convexity condition stated after \re{eq:quad-en} is not innocuous here. Positive semidefiniteness of \re{eq:Bmatrix} is equivalent to $b_1, b_2 \ge 0$ and $\det \ttensor{B} \ge 0$. For the modified Newtonian case \re{ex_modNewton}, with $b_1 = 0$, $b_{12} = \frac{C}{\qrho}$ and $b_2 = \frac{1}{4\pi G}$, one finds
\begin{align}
    \det \ttensor{B} = - \frac{C^2}{\qrho^2} < 0
\end{align}
for every $C \ne 0$: the gradient block is indefinite indicating material instabilities, like in case of classical phase transitions. One can understand the physical implications and relations to dynamic stability in a more general framework.

For the generalized Schr\"odinger--Newton case, presented in details in the next section where in addition $ b_1 = \frac{\hbar^2}{4 m^2 \qrho}$,
\begin{align}
    \det \ttensor{B} = \frac{\hbar^2}{16 \pi G m^2 \qrho} - \frac{C^2}{\qrho^2},
\end{align}
so that positive semidefiniteness holds if and only if
\begin{align}
    \label{eq:densthreshold}
    \qrho \ge \qrho_{\rm c} = \frac{16 \pi G m^2 C^2}{\hbar^2}.
\end{align}

The interpretation of this result requires care, because two distinct instability mechanisms are present and they act at opposite ends of the spectrum. Linearizing \re{eq:sgK-1}--\re{eq:sgK-2} about a homogeneous state $\qrho_0$ shows that the short-wavelength asymptotics of the dispersion relation is governed by the gradient block: 
\begin{align}
    \label{eq:k4asympt}
    \omega^2 \simeq 4\pi G \qrho_0 \det \ttensor{B}(\qrho_0) k^4,
\end{align}
so that $\det\ttensor{B} \ge 0$ is exactly the condition that short wavelengths be
stable, and $\det\ttensor{B} < 0$ gives a growth rate unbounded as $k\to\infty$, i.e.
an initial-value problem that is ill posed in the sense of Hadamard. Convexity of the
gradient block is thus a short-wavelength well-posedness condition. The Jeans
instability, by contrast, resides in the $k$-independent contribution $-4\pi
G\qrho_0$ to $\omega^2$, which originates in the interaction term $a_1(\qrho)\qphi$,
lies outside the gradient block, and is present for every $C$, including $C=0$, where
$\det\ttensor{B} = b_1 b_2 \ge 0$ and the convexity condition is satisfied. The two
mechanisms are therefore independent.

We add that a complete characterization of the linear spectrum of the coupled
matter--field system, including quantum Jeans scales is beyond the scope of the present work.

Finally, two properties of the modified field equation \re{eq:mPoisson} deserve mention. The effective source contains the term $\nabla \cdot \left(\frac{\nabla \qrho}{\qrho}\right)$, whose volume integral reduces to a surface flux of $\frac{\nabla\qrho}{\qrho}$; the integrated gravitational source is therefore unchanged whenever that flux vanishes on the boundary, in particular for a density that becomes uniform at large distances. Locally, however, the effective source, the gravitating density, may be of either sign. }

\subsection{Generalized Schrödinger--Newton equations}

In the light of our previous investigations it is straightforward to formulate even more coupled equations and check their thermodynamic consistency. Assuming the parameters of \re{eq:quad-en} to be $ a_1 = \qrho $, $ a_2 = 0 $, $ b_{1} = \frac{\hbar^2}{4 m^2 \qrho} $, $ b_{12} = \frac{C}{\qrho} $ and $ b_{2} = \frac{1}{4\pi G} $ the \adden \ function \re{ex_SchNewton} results in a generalized self-gravitating Korteweg-like fluid model. The corresponding hydrodynamic equations are
\begin{align}
	\label{eq:sgK-1}
	\dot \qqv &= - \nabla \left( \mu \qf + \phi - \frac{\hbar^2}{2 m^2} \frac{\Delta \sqrt{\qrho}}{\sqrt{\qrho}} - \frac{C}{\qrho} \Delta \phi \right), \\
	\label{eq:sgK-2}
	\Delta \phi &= 4 \pi G \left(\qrho - C\nabla\cdot\left(\frac{\nabla\qrho}{\qrho}\right)\right).
\end{align}

If one does not consider the cross-effects, \ie $ C = 0 $, chemical potential (and correspondingly the fluid pressure) is zero, density is interpreted (and therefore rescaled) as the probabilistic mass distribution $ \qrho = m | \Psi |^2 $ and the external interactions are considered via the potential energy function, then the classical Schrödinger--Newton equation is obtained
\begin{align}
	\label{eq:cSN-1}
	\ii \hbar \pdt \Psi &= \left( - \frac{\hbar^2}{2 m} \Delta + V + m \phi \right) \Psi, \\
	\label{eq:cSN-2}
	\Delta \phi &= 4 \pi G m |\Psi |^2
\end{align}
is obtained. (Let us note here that the system \re{eq:cSN-1}--\re{eq:cSN-2} is sometimes referred to as Schrödinger--Poisson system, and, in this sense the Schrödinger--Newton equation is obtained via substituting the solution of the linear Poisson equation \re{eq:cSN-2} into \re{eq:cSN-1}, \cite{RufBon69a,Dio84a}.) Our thermodynamic methodology proposes a direct generalization for the Schrödinger--Newton equation, when $ C \neq 0 $ and $ \mu \qf \neq 0 $, then the wave function formulation of \re{eq:sgK-1} and \re{eq:sgK-2} are
\begin{align}
	\label{eq:SchN}
	\ii \hbar \pdt \Psi &= \left( - \frac{\hbar^2}{2 m} \Delta + V + m \phi + m \mu \qf - \frac{C}{|\Psi|^2}\Delta \phi \right) \Psi, \\
	\label{eq:NSch}
	\Delta \phi &= 4 \pi G m \left(|\Psi |^2 - \frac{C}{m} \Delta \ln|\Psi|^2 \right).
\end{align}
Let us remark here that in this system of equations seemingly the continuum and particle perspectives are mixing (as presented by the mass multipliers). {However, when the coupling coefficient $ C $ is proportional to the mass of the particle, this mixed picture clears up. This choice seems to be natural for self-gravitating quantum fluids. 

 The dimensions of the coupling constant should be recorded, since they qualify the preceding remark. From the term $\frac{C}{\qrho}\nabla\qrho\cdot\nabla\phi$ in \re{ex_SchNewton}, $C$ has the dimension of mass per length, so the statement that $C$ is proportional to the particle mass is incomplete: it requires a length scale, $C = \frac{m}{\ell}$ for some characteristic length $\ell$. With this identification, and writing $d = \left(\frac{m}{\qrho}\right)^{\nicefrac{1}{3}}$ for the mean
interparticle distance, condition \re{eq:densthreshold} takes the form
\begin{align}
    \ell \ge \frac{d^3}{\ell_{\rm J}^2} = \frac{4\sqrt{\pi G}\, m^{\nicefrac{3}{2}}}{\hbar}\, d^{\nicefrac{3}{2}},
\end{align}
equivalently $\ell\, \ell_{\rm J}^2 \ge d^3$, a dimensionally consistent relation
between the coupling length $\ell$, the quantum Jeans length $\ell_{\rm J}$ and the
mean interparticle distance. We stress that this is a constraint on the admissible
values of the coupling constant at a given density, and not a stability criterion for
the coupled dynamics: as shown in Sec.~\ref{sec:spec}\,D it controls the
short-wavelength well-posedness of the model, while the long-wavelength Jeans
instability is present independently of it.
 The recovery of \re{eq:SchN}--\re{eq:NSch} is moreover conditional on the assumptions collected in Sec.~\ref{app:com-rep}---smooth positive density, irrotationality and single-valued phase---and, for a probabilistic reading of $\qrho = m|\Psi|^2$, on normalization. We therefore present the generalized system as thermodynamically admissible within the stated density range rather than as an unrestricted extension.} 
Finally, let us note that the hydrodynamic formulation of the Schrödinger--Newton equation is particularly useful to investigate its mathematical properties, \eg the existence and uniqueness of solutions under various initial and boundary conditions \cite{AntEta23a,DonEta15a}.

{
Let us remark that if convexity of the quadratic gradient block in \re{eq:quad-en} is retained as an additional stability condition, a particular form of thermodynamic stability. Then positive semidefiniteness implies the density restriction \re{eq:densthreshold}.
Consequently, imposing this additional convexity condition restricts the applicability of the present model to sufficiently large densities. The implications of this restriction, particularly in the low-density limit, and the stability of the complete constrained matter--field system remain to be investigated.
}

{
Independently of whether the additional convexity condition is imposed, the $ C \neq 0 $ wave-function formulation is restricted to regions of strictly positive density, since the terms containing $ \frac{1}{\qrho} $, $ \frac{1}{|\Psi|^2} $, and $ \ln |\Psi|^2 $ become singular as $ \qrho = m |\Psi|^2 $ approaches zero. Consequently, the present formulation does not cover vacuum regions or zeros of the wave function. Its extension to such regions, together with the treatment of global phase and topology, multiply valued velocity potentials, and quantized circulation, provides a natural direction for future investigation. The resulting Schrödinger-like representation should therefore be understood as a local hydrodynamic rewriting of the continuum equations under the assumptions stated above; a physical quantum-mechanical interpretation requires additional assumptions that are not derived in the present work.
}

\subsection{Beyond the quasi-linear quadratic extra energy -- self-consistent self-interacting Newtonian gravity}

If internal variable dependence of the coefficients are also allowed in \re{eq:quad-en}, then one may also gain insight into a broader spectrum of models found in the literature. The motivation for this particular extension is the following. Allowing $b_2$ to depend on the internal variable is the minimal departure from the quasi-quadratic class \re{eq:quad-en}, and the specific choice $b_2 = \frac{c^2}{4 \pi G \phi}$ is the one for which the resulting field equation closes in the form of self-consistent self-interacting Newtonian gravity, a model in which the gravitational field acts as its own source and which has been arrived at independently by other routes~\cite{Ein912a,Giu97a,Fra15a}. Therefore, the previous derivation is repeated for this case. Throughout this subsection we adopt the convention $\phi > 0$, which is required for the transformation to \re{eq_selfg_field} to be well defined; this is legitimate here because in the self-interacting case $\phi$ plays the role of a field amplitude rather than of a potential energy per unit mass, and the requirement fixes the additive constant that is otherwise free in the definition of the gravitational potential.

Let us consider the case of Newtonian gravity, however, the gravitational potential dependence of the coefficient $ b_2 $ is allowed, to be more specific, $ b_2 = \frac{c^2}{4 \pi G \phi} $ with the speed of light $ c $, hence
\begin{align}\label{fu:New-gr}
	\he_{\rm scNG} &= \qrho \phi + \frac{c^2}{8 \pi G} \frac{\nabla \phi \cdot \nabla \phi}{\phi}, \\
	\qqP_{\rm scNG} &= \left( p \qf - \frac{c^2}{8 \pi G} \frac{\nabla \phi \cdot \nabla \phi}{\phi} \right) \ident + \frac{c^2}{4 \pi G} \frac{\nabla \phi \otimes \nabla \phi}{\phi}, \\
	\Phi_{\rm scNG} &= \mu \qf + \phi.
\end{align}

Let us draw the attention on the \adden \ density, where a term similar to the Fisher energy \re{eq:Fisher} appears, which in quantum mechanics leads to the Bohm potential. Although the pressure tensor has been modified, the corresponding holographic potential remains the same as in the case of classical Newtonian gravity. In this case, the field equation corresponds to self-consistent self-interacting Newtonian gravity \cite{Ein912a,Giu97a,Fra15a}, \ie
\begin{align}
	\label{eq_selfg_field}
	\Delta\phi = \frac{4\pi G}{c^2} \left( \qrho \phi +\frac{c^2}{8 \pi G} \frac{\nabla\phi\cdot\nabla\phi}{\phi} \right),
\end{align}	
which can be formulated in a Bohmian form as
\begin{align}
	\frac{\Delta \sqrt{\phi}}{\sqrt{\phi}} = \frac{2 \pi G}{c^2} \qrho.
\end{align}
{The significance of this form is structural rather than interpretational. The field equation for $\sqrt{\phi}$ has exactly the shape of the Bohm-potential relation \re{eq:Fisher} of the Korteweg case, with $\sqrt{\phi}$ in place of $\sqrt{\qrho}$: the same Fisher-information-type gradient structure that produces the quantum potential also produces the self-interaction of the gravitational field. This is a further instance of the unifying structure that is the subject of this paper, and we note that it carries no implication that the gravitational field is thereby quantum.
}

\section{Conclusions and outlook}\label{sec:disc}

\paragraph*{Plain-language summary.}
We have shown that the equations governing several seemingly distinct continuum theories---ordinary fluids, capillary fluids, the hydrodynamic form of the Schr\"odinger equation, Newtonian self-gravitating gas clouds, and the Schr\"odinger--Newton system that couples quantum matter to its own gravitational field---can all be derived {within a common framework by applying the Second Law of Thermodynamics under the constitutive and modeling assumptions stated in this work}. The key common feature, which we call \emph{classical holography}, is that {under the identified conditions, for non-dissipative fluids in this family,} the divergence of the pressure tensor is the gradient of a scalar potential, so that field equations and particle-like equations of motion become two faces of the same coin. A direct consequence is that, {for sufficiently regular flows} in the absence of rotation, the fluid description can be repackaged into a single complex equation of Schr\"odinger type. The framework also identifies a previously unnoticed density--gravity coupling whose experimental and astrophysical consequences are open questions for future work.%
	
	\bigskip
	
	The general unification of gravity and quantum mechanics is a long-lasting issue in physics; \eg the Schr\"odinger--Newton equation can unify certain aspects. In this work{,} we have presented a methodology based on nonequilibrium thermodynamics and hydrodynamics{,} which{,} as a particular example, results in the Schr\"odinger--Newton equation. Although within the framework of classical physics, this is an unexpected result{,} and our work may propose further developments in this field.
	
	Nonequilibrium thermodynamics provides a unified framework for continuum theories. The only requirement of the field is that the dissipation inequality must be satisfied, hence a sufficiently general thermodynamic approach is independent{} and{,} therefore, compatible with any microscopic or mesoscopic background. Evaluation of the dissipation inequality results in the most general, but thermodynamically compatible forms of the constitutive functions {within the chosen constitutive state space and under the stated assumptions}. Within this framework, we have presented a generalized, weakly nonlocal fluid model whose dynamics also involve an internal variable.
	
	Although nonequilibrium thermodynamics is known as the science of dissipation and irreversibility, it proved to be constructive also in the analysis of perfect (\ie dissipation{-}free) continua, with unexpected {general consequences}, such as {the} classical holographic property and the functional derivatives emerging without any variational principles. According to the holographic property, {the} divergence of the pressure tensor is equivalently formulated via the gradient of a scalar potential. {We have also identified the precise thermodynamic condition under which this holds, $\nabla s \times \nabla T = \zero$, and shown that the weakly nonlocal sector contributes no obstruction to it, so that the only obstruction is the classical thermal one already present for an Euler fluid. The residual non-potential force is the continuum entropic force $T\nabla s$.} An interesting further implication
	 {of the holographic momentum balance is a vorticity transport law that preserves sufficiently regular, initially irrotational flows and enables} the wave-function representation of the dynamic equations. Thus, {a continuum-mechanical route to the hydrodynamic representation of certain quantum models is established}.
	
	We have formulated some important models in the framework of perfect fluids by assuming a quadratic energy density, from which the reversible part of the pressure tensor and the corresponding holographic potential can be derived. The governing equations of Korteweg-like fluids (including Schr\"odinger--Madelung fluids), self-gravitating classical fluids and self-gravitating quantum fluids (\ie the Schr\"odinger--Newton equation) are reproduced, and beyond these models, we have recognized a possible density--gravity coupling energy term{} as an example of further generalizations. These models show that the modification of the field equation alone may be inconsistent without the complete theory. The field equation is derived from the energy density, which determines the pressure tensor{,} which leads to the equation of motion{,} and vice versa.
    	
	Naturally, there are several conceptual issues and questions that should be addressed when the presented thermodynamic approach is compared to our {current} understanding of physics.
	Here, we have collected the most important ones:
	\begin{itemize}
		\item For non-relativistic continua, there are only a few attempts to improve the usual relative, reference-frame-dependent framework. However, a rigorous reference-frame-independent treatment using four-quantities may reveal further consequences and restrictions. For instance, classical holography seems to be a property of the energy-momentum tensor, sometimes fused with the energy-momentum balance. After such an analysis, it is worth generalizing our study to the special-relativistic spacetime model.
		\item In Newtonian gravity (and also in electromagnetism, see \eg Ref.~\cite{Mat24a}) the energy densities are often discussed. A family of different expressions---usually through Lagrangian densities---are recommended, which all result in the same integral energy~\cite{Pet81a}. Our expression $\he_{\rm N} = \qrho \phi + \tfrac{1}{2}\tfrac{1}{4\pi G}\nabla\phi\cdot\nabla\phi$, the Ohanian gravitational energy density, seems to be distinguished from {a} thermodynamic point of view, but its connection to classical field theories should be investigated. This question is analyzed in detail (work in progress).
		\item Regarding holography, in our formulation{,} a dissipation-free cross-effect appears through the antisymmetric coupling between the scalar thermodynamic fluxes [see \re{eq:Ons-1} and \re{eq:Ons-2}], which may have interesting consequences for the dynamics and the formulation of the holographic potential. We plan to analyze this possibility in detail in the future.
	\end{itemize}
	
	\section*{Acknowledgments}
	
	This work was supported by a grant from the National Research, Development and Innovation Office (TKKP Advanced 150038). The authors acknowledge networking support by the grant NKFIH NKKP-Advanced 150038 and STSM Grant from FuSe COST Action [CA24101], funded by COST (European Cooperation in Science and Technology). Project no.~TKP-6-6/PALY-2021 has been implemented with the support provided by the Ministry of Culture and Innovation of Hungary from the National Research, Development and Innovation Fund, financed under the TKP2021-NVA funding scheme. The authors thank Roberto Trasarti-Battistoni, Tam\'as F\"ul\"op, R\'obert Kov\'acs, and M\'at\'e Pszota for valuable discussions. {The authors highly appreciate the constructive comments and suggestions from both reviewers.}
	
	\section*{Author Declarations}
	
	\subsection*{Conflict of Interest}
	The authors have no conflicts to disclose.
	
	\subsection*{Author Contributions}
	\textbf{M\'aty\'as Sz\"ucs:} Conceptualization (equal); Formal analysis (equal); Methodology (equal); Writing -- original draft (equal); Writing -- review \& editing (equal). \textbf{P\'eter V\'an:} Conceptualization (equal); Formal analysis (equal); Methodology (equal); Supervision (lead); Writing -- original draft (equal); Writing -- review \& editing (equal).
	
	\subsection*{Data Availability}
Data sharing is not applicable to this article as no new data were created or analyzed in this study.
	

%

\end{document}